\documentclass[12pt]{spieman}  
\usepackage{amsmath,amsfonts,amssymb}
\usepackage{graphicx}
\usepackage{setspace}
\usepackage{tocloft}
\usepackage[numbers]{natbib}
\usepackage{upgreek}

\usepackage{soul} 
\usepackage{lineno}

\DeclareMathOperator*{\argmin}{arg\,min}

\newcommand{\yp}{y^i_p}
\newcommand{\yf}{y^i_f}
\newcommand{\up}{u^i_p}
\newcommand{\uf}{u^i_f}

\newcommand{\edited}[1]{\textcolor{black}{#1}}

\title{Data-driven subspace predictive control of adaptive optics for high-contrast imaging.}

\author[a]{Sebastiaan Y. Haffert}
\author[a]{Jared R. Males}
\author[a]{Laird M. Close}
\author[a,b,c]{Kyle Van Gorkom}
\author[a]{Joseph D. Long}
\author[a,b]{Alexander D. Hedglen}
\author[a,b,d,e]{Olivier Guyon}
\author[a,b]{Lauren Schatz}
\author[a,b]{Maggie Kautz}
\author[a,b]{Jennifer Lumbres}
\author[a,b]{Alex Rodack}
\author[a,b]{Justin M. Knight}
\author[f]{He Sun}
\author[g,h]{Kevin Fogarty}

\affil[a]{University of Arizona, Steward Observatory, Tucson, Arizona, United States}
\affil[b]{Wyant College of Optical Science, University of Arizona, 1630 E University Blvd, Tucson, AZ 85719, USA}
\affil[c]{NASA Goddard Space Flight Center, Greenbelt, MD 20771, USA}
\affil[d]{Astrobiology Center, National Institutes of Natural Sciences, 2-21-1 Osawa, Mitaka, Tokyo, JAPAN}
\affil[e]{National Astronomical Observatory of Japan, Subaru Telescope, National Institutes of Natural Sciences, Hilo, HI 96720, USA}
\affil[f]{Department of Computing and Mathematical Science, California Institute of Technology, Pasadena, CA 91125, USA}
\affil[g]{The Division of Physics, Mathematics and Astronomy, California Institute of Technology, Pasadena, CA 91125, USA}
\affil[h]{NASA Ames Research Center, Moffett Field, California 94035 USA}

\cftpagenumbersoff{figure}
\cftpagenumbersoff{table} 
\begin{document} 
\maketitle

\begin{abstract}
  The search for exoplanets is pushing adaptive optics systems on ground-based telescopes to their limits. One of the major limitations at small angular separations, exactly where exoplanets are predicted to be, is the servo-lag of the adaptive optics systems. The servo-lag error can be \edited{reduced} with predictive control where the control is based on the future state of the atmospheric disturbance. We propose to use a linear data-driven integral predictive controller based on subspace methods that is updated in real time. The new controller only uses the measured wavefront errors and the changes in the deformable mirror commands, which allows for closed-loop operation without requiring pseudo-open loop reconstruction. This enables operation with non-linear wavefront sensors such as the pyramid wavefront sensor. We show that the proposed controller performs near-optimal control in simulations for both stationary and non-stationary disturbances and that we are able to gain several orders of magnitude in raw contrast. The algorithm has been demonstrated in the lab with MagAO-X, where we gain more than two orders of magnitude in contrast.
\end{abstract}


{\noindent \footnotesize\textbf{*} NASA Hubble Fellow, \linkable{shaffert@arizona.edu} }

\begin{spacing}{2}   

\section{Introduction}
\label{sect:intro}  
The upcoming generation of Giant Segmented Mirror Telescopes (GSMT) have the light gathering capability and angular resolution to directly image Earth-like planets around other stars, which allow us to search for bio-signatures. However, these ground-based telescopes do not operate at their diffraction-limit due to turbulence in the Earth's atmosphere. And even if these large telescopes could be used close to their diffraction-limit, the star is usually orders of magnitude brighter than the planet, making it difficult to distinguish the planet from the star\cite{traub2010exoplanets}. High-contrast imaging (HCI) instruments are designed to overcome these challenges by using extreme adaptive optics (ExAO) to compensate for atmospheric disturbances and recover the angular resolution, while advanced coronagraphs are used to remove the influence of starlight\cite{guyon2018exao}.

With the current generation of HCI instruments \cite{macintosh2014gpi,beuzit2019sphere,jovanovic2015scexao} we can routinely reach post-processed contrast levels of $10^{-4}$ to $10^{-6}$, depending on the angular distance from the host star. With these contrast levels, we are sensitive to hot and massive self-luminous planets\cite{marley2007hotjupiters}. Even though we are sensitive to massive Jupiter-like planets, only a few planets have been imaged and spectroscopically characterized. The results from large surveys that targeted massive Jupiters on wide orbits indicate that the planet occurrence rate drops sharply between 1 to 10 AU \cite{bowler2015gpoccurence, nielsen2019gpies, fernandes2019occurence, wagner2019wideoccurence}. 

More planets could be directly imaged if the sensitivity close to the star is improved. The main limiting factor close to the star is the performance of the adaptive optics (AO) systems\cite{kasper2012hci,milli2017sphereperformance,cantalloube2019winddrivenhalo}. During the operation of an AO system a wavefront sensor measures the wavefront aberrations, which are then fed back to a deformable mirror for correction. This causes an inevitable delay because the system can only compensate after it has measured the disturbance. There are two options to \edited{reduce} the servo-lag error in AO systems, either run the full system  at a higher speed, which is currently being tried \cite{males2018magaox}, or by the use of predictive control. \edited{In predictive control a model of the system is used to make a prediction about the future state of the system. This allows the predictive controller to anticipate future behavior and mitigate errors before they happen.} Predictive control has been proposed some time ago as solution to the servo-lag error\cite{gavel2003optstrehl}. \edited{If predictive control is successfully implemented it could lead to a gain of 2 or 3 orders of magnitude in contrast, under the assumption that much of the temporal evolution is predictable \cite{guyon2017eof,males2018lpc,correia2020hcipwfs}.} 

 Initial results from predictive control focused on low-order modes, which was mainly due to restrictions in computing power and the availability of high-order deformable mirrors (DM) \cite{dessenne1997modalcontrol,dessenne1999onsky}. From that point two distinct paths of predictive control appeared, one focused on physical modelling of the atmosphere and its dynamics \cite{poyneer2007fpc,poyneer2010pfc,kulcsar2006LQG} and the other focused on data-driven methods \cite{hinnen2005ssc,guyon2017eof}. Recently a large amount of effort has been put into investigating the application of deep reinforcement learning for predictive control\cite{xu2019deep, gomez2019experience}. There are also algorithms where model-based and data-driven approaches are merged \cite{ sivo2014canary}.
 
 Currently, both paths are maturing and are being implemented in laboratory and on-sky demonstrations. The initial results are encouraging, showing increased Strehl ratio in many conditions \cite{sivo2014canary,tesch2015palomar}. The gain in contrast for ExAO systems however has been limited \cite{jensenclem2019keck}. A major limitation for ExAO systems that is common in the previous methods is that they require a sequence of open-loop wavefront sensor measurements to estimate certain aspects of the system, e.g the Power Spectral Density (PSD) \cite{dessenne1999onsky}, \edited{correlation matrix} of all inputs and outputs \cite{guyon2017eof}, or the wind velocity of turbulent layers \cite{poyneer2007fpc}. The wavefront sensor of interest for HCI is the Pyramid Wavefront Sensor \cite{ragazzoni1996pwfs}, which has limited dynamic range. Due to the limited dynamic range it may not be possible to reconstruct the disturbance PSD from open-loop measurements. \edited{Furthermore, variable calibration and inherent non-linearity of the PWFS can change the optical gain, which effectively changes the modal sensitivity } \cite{deo2019opticalgaincalibration}. \edited{If there is no correction for the optical gain, the pseudo-open loop reconstruction will under or overestimate the wavefront. Online gain tracking is necessary to compensate for this effect. And finally, the atmosphere itself is dynamic and exhibits non-stationary turbulence. This is limiting the application of open-loop predictive controllers \cite{kooten2019wind}.}
 
 Another complication stems from the use of multi-stage AO systems, where several dynamical components are operated at the same time, but each may be operated at different frequencies. The temporal dynamics of each component need to be accurately accounted for to reconstruct the full open-loop disturbance. Any model error in the DM dynamics immediately folds into the disturbance reconstruction. Additionally, ExAO systems are being operated at very high speeds close to the operating frequency of DMs, which means that the DM can not be modelled with a fixed-lag step response and the temporally resolved dynamics need to be taken into account. 

In this work, we propose a completely model-free data-driven method that identifies the atmospheric and system dynamics in closed-loop, and is updated online to track non-stationarity. Our method is based on a closed-loop data-driven subspace identification algorithm, which has been gaining traction recently \cite{favoreel1999mflqg,huang2008dynamic}. The principle of the Data-Driven closed-loop Subspace Predictive Controller (DDSPC) is to obtain the predictive controller directly from the measured input-output pairs during closed-loop operations, without any intermediate parametric model identification. In this work, we will derive a predictive controller based on the DDSPC method that uses integral-action, which will drastically lower the influence of model errors and allows it to be used in closed-loop.

In Section 2, we derive the closed-loop data-driven subspace identification algorithm and its controller. We numerically demonstrate the power of the algorithm for predictive control in Section 3, and in Section 4 we show the verification of the algorithm in a lab setting. Section 5 concludes the manuscript and gives an outlook on the implementation for on-sky.

\section{Data-driven closed-loop subspace identification}
The proposed DDSPC controller works in a mode space of the DM, e.g. actuator space or Zernike space. We assume that the modal coefficients are retrieved from a WFS that measures the wavefront error in closed-loop at regular time intervals. The reconstructed wavefront error at timestep $i$ is then defined as $\Delta y_i$. At each time step $i$ there is also a new command that will be sent to the DM to compensate for the wavefront errors, which is defined as $\Delta u_i$. It is important to note here that the DDSPC controller is working with the changes in the wavefront and not the full wavefront and actuator commands. This will allow the predictive controller to retain integral action. 

\subsection{Defining the auto-regressive model}
The key aspect of the predictive controller lays in its ability to predict the future WFS measurements. The future measurements of the WFS depend on the past wavefront sensor measurements from the atmosphere changing between time steps and past DM commands due to non-instantaneous DM dynamics. And finally the future measurements also depend on the future commands, because at every time step a new command will be send to the DM which has temporal dynamics that need to be taken into account. We will assume that a linear auto-regressive (AR) model is enough to capture all dynamics of the system. The AR model for the measurement at the next time step is,
\begin{equation}
\label{eq:ARmodel}
\Delta y_{i+1} = \sum_{n=i-N}^{n=i} a_n \Delta y_{n} + \sum_{n=i-N}^{n=i} b_n \Delta u_{n} + \sum_{n=i+1}^{n=i+M} c_n \Delta u_{n}.
\end{equation}
This equation consists of three parts, the AR model for the past $N$ measurements with coefficient $a_n$, the AR model for the past $N$ commands  with coefficients $b_n$, and finally the AR model for the $M$ future commands with coefficients $c_n$. \edited{The order ($N$) for the past commands and measurements do not have to be equal. For practical purposes we kept them equal.} The model of Eq. \ref{eq:ARmodel} can be rewritten in a clearer vector notation by defining the data vector which contains a sequence of length $N$ of either measurements or commands as,
\begin{equation}
    x_{i+N:i} = \begin{pmatrix} x_{i+N}& \hdots &x_{i}  \end{pmatrix}^T.
\end{equation}
Here $x_i$ is the value of variable $x$ at time step $i$. If a prediction horizon of $M$ steps is considered with $N$ past measurements, the notation can be further simplified. To do this, the future measurements at time step $i$ are defined as $\yf= \Delta y_{i+M:i+1}$ and the past measurements as $\yp = \Delta y_{i:i-N}$. The future and past commands are defined in a similar manner, $\uf=\Delta u_{i+M:i+1}$ and $\up=\Delta u_{i:i-N}$. With the simplified notation the AR model becomes,
\begin{equation}
    \yf = A \yp + B \up + C \uf.
    \label{eq:prediction}
\end{equation}
With $A$ the prediction matrix of size $M \times N$ that correlates the past $N$ measurements to the future $M$ measurements, $B$ the prediction matrix of size $M \times N$ that correlates the past $N$ commands to the future $M$ measurements, and $C$ the prediction matrix of size $M \times M$ that correlates the future commands to the future measurements.

\subsection{Online estimation of the prediction matrices}
The AR model needs to be updated every time step to take into account non-stationary statistics. Therefore, we need to do an online identification of the AR model. The problem can be reduced to a single matrix-vector multiplication because it is a linear problem,
\begin{equation}
y^i_f = \begin{bmatrix} A^i&  B^i&  C^i& \end{bmatrix} \begin{bmatrix} y^i_p\\  u^i_p\\  u^i_f \end{bmatrix} = \Theta^i \phi^i
\end{equation}
Here $\phi$ is the concatenation of $\yp$, $\up$ and $\uf$, while $\Theta$ is the concatenation of $A$, $B$ and $C$. This can be solved in a straight forward way by defining the least-squares solution,
\begin{equation}
\Theta = \argmin_{\Theta} \left \| y^i_f - \Theta \phi^i  \right \|_i.
\end{equation}

Recursive least-squares (RLS) is used to update our prediction matrix at every time step. RLS consists of 4 steps, of which the first is the determination of the update gain,
\begin{equation}
    K^{i} = \frac{\gamma^{-1}\phi^{i}P^{i-1} }{1 + \gamma^{-1} \phi^{iT}P^{i-1}\phi^i}.
\end{equation}
Here $K^i$ is the gain matrix at time step $i$, $\gamma$ is the forgetting factor that is included to reduce the weight of measurements far in the past, and $P^i$ is the inverse covariance matrix between all features at time step $i$. The next step is to calculate the prediction error at step $i$,
\begin{equation}
    e^i = \yf - \Theta\phi^i.
\end{equation}
After which the prediction matrix can be updated,
\begin{equation}
    \Theta^{i+1} = \Theta^i + K^ie^i.
\end{equation}
And finally the inverse covariance matrix is updated
\begin{equation}
    P^{i} = \gamma^{-1}P^{i-1} - \gamma^{-1}K^i\phi^iP^{i-1}.
\end{equation}
The RLS algorithm is closely related to the Kalman Filter. A benefit of the RLS algorithm is that it is an $\mathcal{O}(n^2)$ algorithm as opposed to the Kalman Filter which is an $\mathcal{O}(n^3)$ algorithm. This difference in algorithmic complexity allows the RLS algorithm to update its matrices in real time, while for the Kalman Filter the gain matrix is almost always calculated offline \cite{correia2017dkf}. 


\subsection{Derivation of the controller}
The results from the previous two sections can now be used to derive the finite horizon least-squares optimal controller. First, a cost function has to be defined that the controller will optimize. A logical choice is the mean square error of the future $M$ wavefront errors because those are the errors that are predicted. This leads to the following cost function,
\begin{equation}
    J_i = \sum_{k=1}^{M} y_{i+k}^T y_{i+k} + \lambda \sum_{k=1}^{M} u_{i+k}^T u_{i+k} = y_f^{iT}y_f^{i} + \lambda u_f^{iT}u_f^{i}.
    \label{eq:cost}
\end{equation}
The future commands have been added to the cost function with a scalar $\lambda$ to penalize large commands, which effectively regularizes the control algorithm. The best estimate that is available at time step $i$ of the future measurements is the AR model described in Eq. \ref{eq:prediction}, which upon substituting into Eq. \ref{eq:cost} will provide a relation between the past measurements and control signals and the future control signals.

\begin{equation}
    J_i = y_f^{iT}y_f^{i} + \lambda u_f^{iT}u_f^{i} = \begin{bmatrix}
y_p^{iT} & u_p^{iT} & u_f^{iT}
\end{bmatrix} \begin{bmatrix}
A^TA & A^TB & A^TC \\ 
B^TA & B^TB & B^TC\\ 
C^TA & C^TB & C^TC
\end{bmatrix} 
\begin{bmatrix}
y_p^i \\
u_p^i \\
u_f^i
\end{bmatrix} 
+ \lambda u_f^{iT} u_f^i.
\end{equation}

The regularisation $\lambda$ can be absorbed into the cost matrix,
\begin{equation}
    J_i = \begin{bmatrix}
y_p^{iT} & u_p^{iT} & u_f^{iT}
\end{bmatrix} \begin{bmatrix}
A^TA & A^TB & A^TC \\ 
B^TA & B^TB & B^TC\\ 
C^TA & C^TB & C^TC + \lambda I
\end{bmatrix} 
\begin{bmatrix}
y_p^i \\
u_p^i \\
u_f^i
\end{bmatrix} 
\end{equation}
The optimal controller can now be derived by finding the future control commands that minimize the cost function
\begin{equation}
    u_f^i = \argmin_{u_f^i} J_i.
\end{equation}
This is relatively straightforward because the cost function is quadratic. To find the minimum, we take the derivative of the cost function with respect to the future commands and equate it to zero,
\begin{equation}
    \frac{\partial J_i}{\partial u_f^i} = 2 \begin{bmatrix}
A^TC & B^TC
\end{bmatrix}
\begin{bmatrix}
y_p^i \\
u_p^i
\end{bmatrix} + 2 \left(C^TC + \lambda I \right) u_f = 0.
\end{equation}
After some algebra we find the optimal control signal as,
\begin{equation}
    u_f = -\left(C^TC + \lambda I\right)^{-1}\begin{bmatrix}
A^TC & B^TC
\end{bmatrix}
\begin{bmatrix}
y_p^i \\
u_p^i
\end{bmatrix}.
    \label{eq:controller}
\end{equation}
The controller itself is $K = -\left(C^TC + \lambda I\right)^{-1}\begin{bmatrix}
A^TC & B^TC
\end{bmatrix}$. This controller calculates the optimal control sequence for the next $M$ steps. However, because the optimal control sequence is recalculated at every time step, only the next command is important.

\subsection{Computational considerations}
Even though the RLS is more efficient than the Kalman filter, it is still a computational challenge to apply the DDSPC with coupling between all modes for a high-order AO system with thousands of modes. To ease the computation time, we implemented a distributed control scheme where we create a controller for each mode. This has the advantage that it is naturally parallel and therefore very easy implement on distributed computing architectures, such as CPU or GPU clusters. A \edited{basis that is close to orthogonal} is required for the distributed control. This will prevent the spatial correlations from folding into temporal correlations, which is otherwise very difficult to correct. The Gaussian actuator basis with low actuator cross talk or the Fourier basis are examples of possible mode bases. In this work, we will use the actuator basis.

The distributed controller works so long as spatio-temporal correlations between modes can be ignored. \edited{Recent work \cite{kooten2020sphere} showed with telemetry from VLT/SPHERE \cite{beuzit2019sphere} that using the past history of a mode is enough to predict the future state. Spatial correlations do not improve the prediction accuracy if the system is driven at high speeds because it takes too long for the wind to move the phase across a sub-aperture. For an AO system like SPHERE the time history needs to be at least 13 steps. For this we assumed a 1kHz loop speed, a windspeed of 15m/s and 41 wfs pixels across the pupil. And this is only the case for single-layer frozen flow turbulence. The influence of spatial correlations will be lowered even more if multiple (possibly boiling) layers are moving across the aperture.} This implies that the proposed distributed approach is valid and we do not have to consider modal correlations for predictive control. Distributed model predictive control also has been shown to work well in simulations \cite{poyneer2007fpc, correia2017dkf}. Later sections in this work will show that the distributed approximation still results in near perfect correction in our end-to-end simulations.

\subsection{Parameter investigation and stability}
\edited{It is not immediately clear that data-driven methods are have guaranteed to be stable. A lot of research has focused on stability of data-driven methods in the past years, of which a lot has been driven by the advances in (deep) reinforced learning. At first closed-loop data-driven model predictive control (MPC) was shown to be have stability guarantees \cite{berberich2019datadrivenstability}. Later the stability was expanded to the data-driven subspace predictive control scheme \cite{coulson2020ddstability}. The DDSPC that is proposed here is slightly different because we use the closed-loop residuals and the input-output model is update online. The data-driven MPC control is guaranteed to be stable if three conditions are met; the cost function needs to have a quadratic upper bound, the number of total number of data samples needs to be larger than the number of free parameters in the model, and the history length $N$ needs to be larger than the future length $M$. While we have not proven that these conditions guarantee a stable controller for the DDSPC, they can still be used as a guideline when choosing the parameters of the controller. It is easy to meet the conditions. The quadratic upper bound is guaranteed because we use a quadratic cost function. The number of data samples puts a constraint on the forgetting factor. Each time a new data point is added the old data will be multiplied by $\gamma$, which means that after $T$ steps the weight of the data will be $\gamma^T$. We now assume that the data does not contribute anymore if its weight is smaller or equal to a threshold value $\epsilon$. So if we want to retain at least $T$ samples the inequality $\gamma^T \geq \epsilon$ must hold. The proposed model has $M\times(2N + M)$ free parameters. Consequently, the constraint on the forgetting factor is then $\gamma \geq \epsilon^{1/(2NM + M^2)}$. A typical model has between 40 and 100 free parameters. With a threshold value of $\epsilon=0.1$, we find a lower bound of about $0.955$ for the small models and a lower bound of $0.975$ for the larger models. In future work we will investigate if these conditions are also sufficient to guarantee stability for the DDSPC algorithm. Here, the stability is analyzed with end-to-end simulations. }

\begin{figure}
    \begin{center}
        \includegraphics{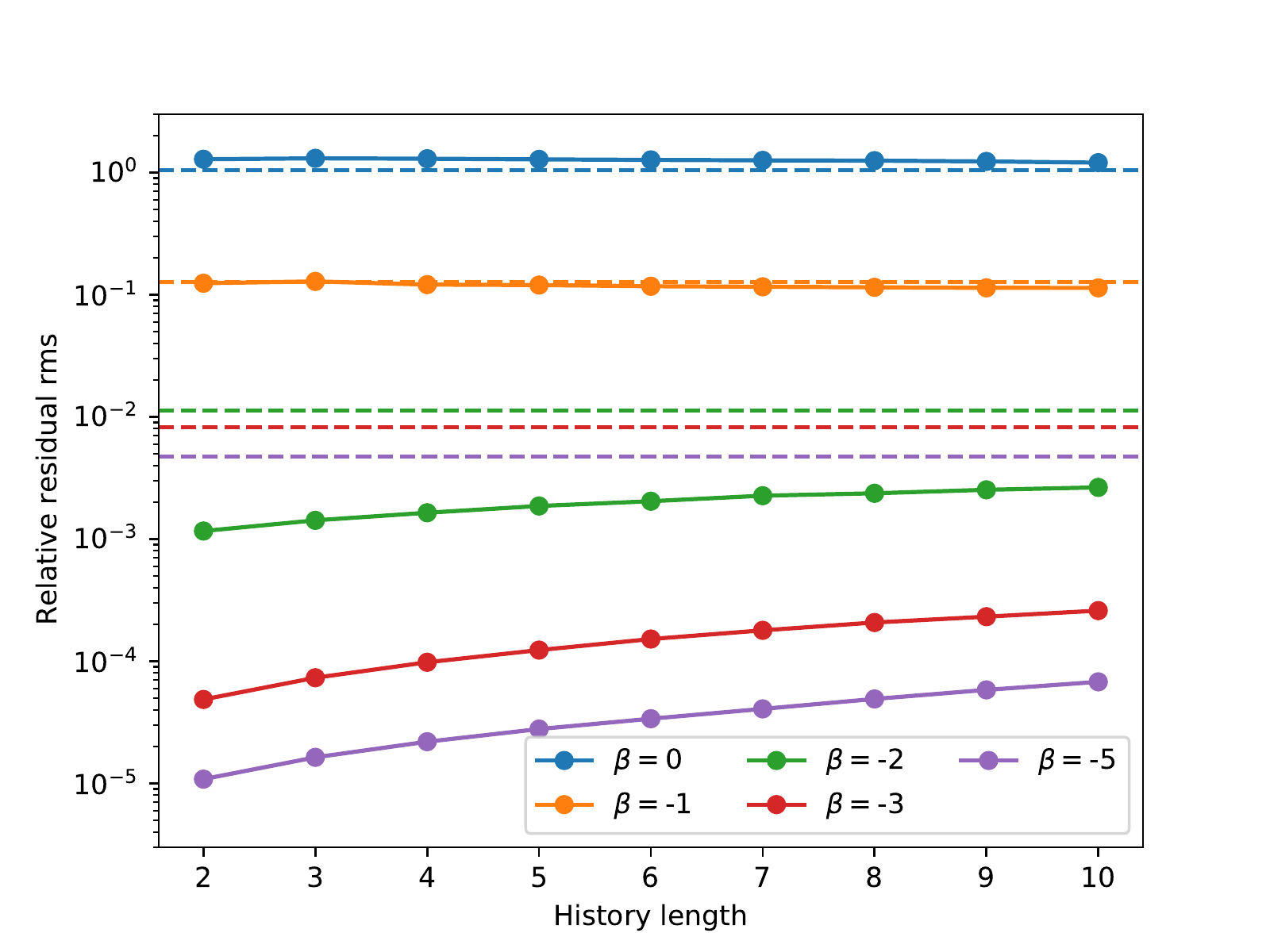}
	\end{center}
    \caption{The standard deviation of the residuals during closed-loop operations. The different colors indicate different power-law indices. The dashed lines represent the residuals for the gain optimized integrator. The solid lines show the residuals of the DDSPC. For $\beta\leq-2$, the predictive control has residuals that are at least an order of magnitude smaller. The residuals increase as the history length is increased for the steeper power-laws. For the shallow power-law disturbances, $\beta\geq-1$, the integrator and predictor behave quite similar.}
    \label{fig:history} 
\end{figure}

\edited{The stability is tested by applying the DDSPC algorithm with varying history length ($N$) and injecting a range of power-law disturbances. The disturbances have the form of $P\propto f^{\beta}$, with $f$ the temporal frequency and $\beta$ the power-law index. Each generated time-series is normalized to a standard deviation of 1.0. For these simulations we used a single frame delay. The prediction horizon ($M$) was set to 3 for these tests. The relative RMS as function of history length is shown in Figure \ref{fig:history} and compared against a gain optimized integrator. This Figure shows that the DDSPC works significantly better than an integrator for power-law indices $\beta\leq-2$. For flat power spectra ($\beta\geq-1$), the integrator and predictor have about equal performance. Longer history lengths decrease the performance of the controller for the steeper power laws. While the shallow power laws prefer a longer history length that can be used to average out the noise. This also implies that there is an optimal history length when a mixture of shallow and steep power laws are present in the disturbance.}

\begin{figure}
    \begin{center}
        \includegraphics{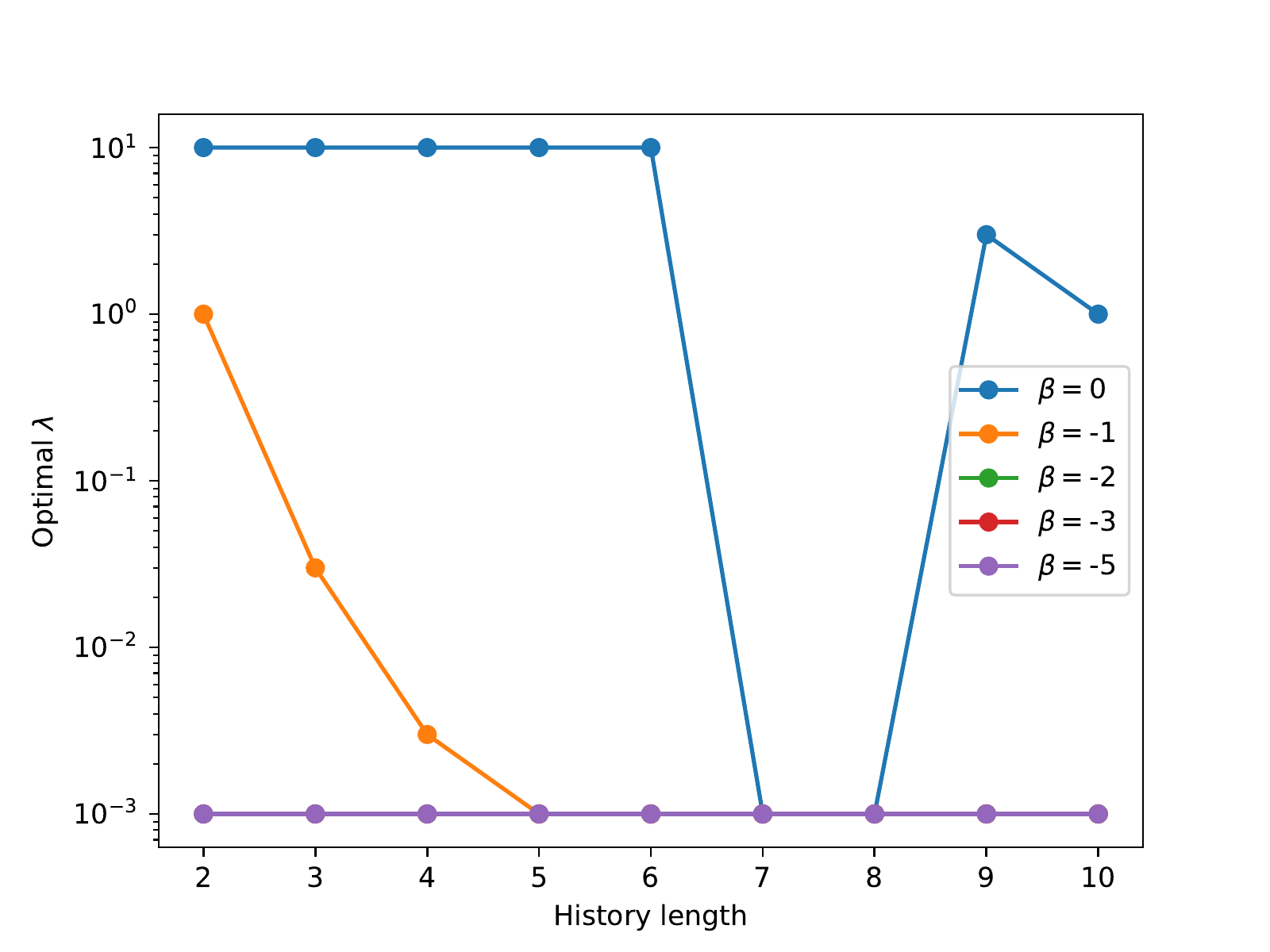}
	\end{center}
    \caption{The optimal regularization parameter for different power-law indices and different history lengths. The different colors represent different power-law indices. The optimal value is the smallest value that was considered for the steep power-laws ($\beta<-2$). For the shallow power-laws a strong regularization is preferred.}
    \label{fig:regularization} 
\end{figure}

\edited{The DDSPC has the regularization parameter $\lambda$ as hyper parameter, which was optimized for each power-law index and each history length. The optimal $\lambda$ is shown in Figure \ref{fig:regularization}, which shows that the optimal regularization parameter should be very small $\lambda=10^{-3}$ for most situations. For the nosiest cases a strong regularization is preferred. This is expected because the optimal control is no control when either the measurements or the disturbance itself consist of pure noise. }

\begin{figure}
    \begin{center}
        \includegraphics{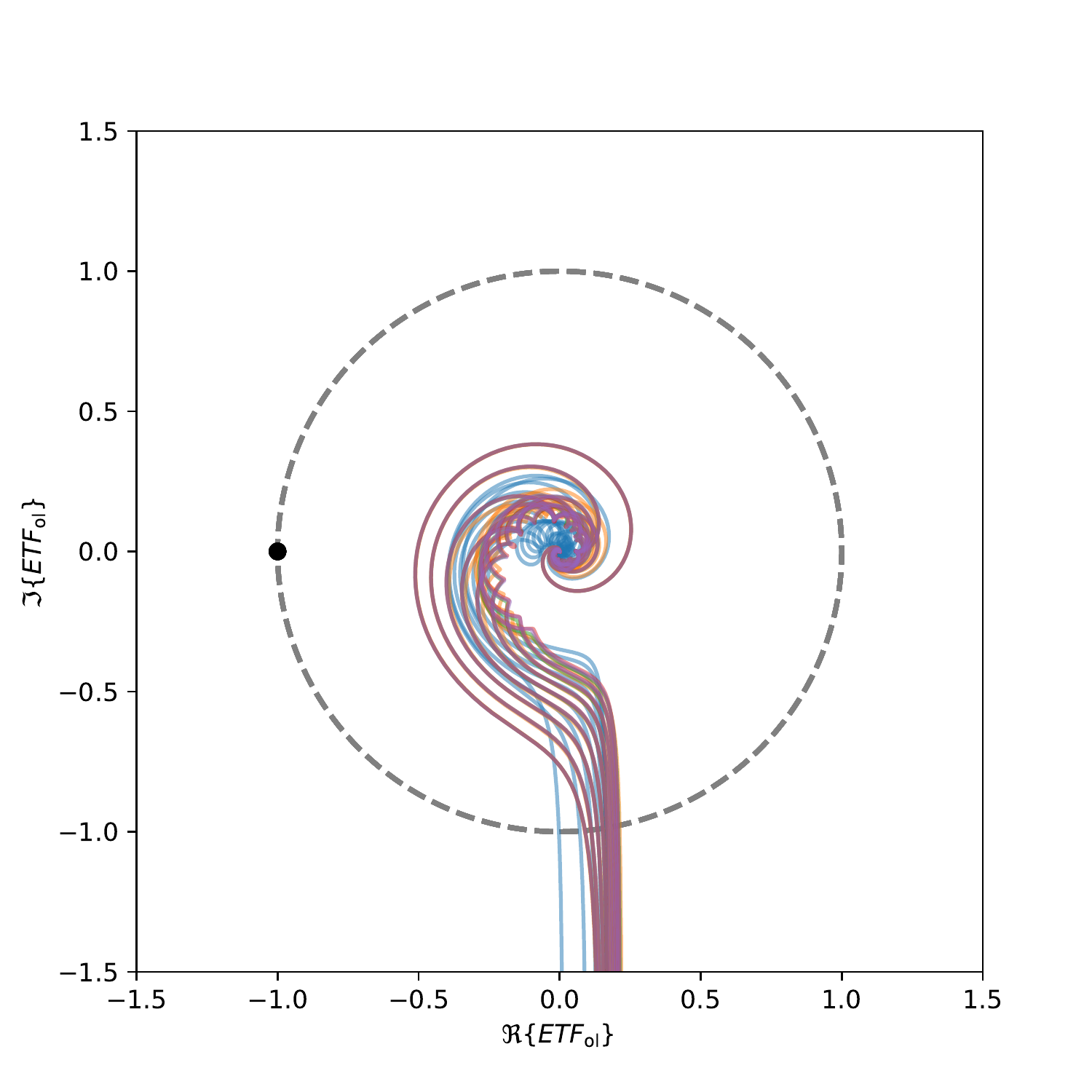}
	\end{center}
    \caption{Nyquist plot demonstrating the stability of the DDSPC controller for the cases considered in this section. The open-loop ETFs are plotted in the imaginary plane. Note that frequency increases as the lines move out from the center. The controllers are stable so long as they do not encircle the black point $(-1, 0)$. The gray dashed circle has a radius of 1. We show the open-loop ETFs for each controller. The color represents different power-law indices. All controllers are stable because they do not encircle the point $(-1, 0)$.}
    \label{fig:stability} 
\end{figure}

\edited{The stability of the controllers is verified by inspecting their Nyquist Plot. The open-loop Error Transfer Function (ETF) is created from the controller ($K$) using the equation for the frequency response of general linear filter\cite{males2018lpc}. The Nyquist plot of each controller is shown in Figure \ref{fig:stability}, which plots the real versus the imaginary part of the open-loop ETF. The system is stable in closed-loop if the ETF does not contain $-1+0i$ as a pole, which can be determined by the number of encirclements of the point $(-1, 0)$. All systems do not encircle the unstable pole, which means that all derived controllers provide stable feedback. From this we determine that the DDSPC is stable over a wide range of disturbances and parameter choices. These results do not guarantee stability for all systems. However, we have not yet encountered a situation, either in simulation or lab experiments, where the controller is unstable. }

\subsection{Learning time}
\label{sec:learning}
\edited{Data-driven algorithms will only have an advantage if they can learn the system parameters fast enough, which makes it important to examine the learning time scale. A long time sequence of 30 seconds is simulated to estimate the scaling behavior of the algorithm. Frozen-flow turbulence with the parameters from \ref{tab:parameters} is used as the input disturbance.} The initial prediction matrix is set to zero and inverse covariance matrix is initialized as a diagonal matrix with a large number, such as $10^9$, on the diagonal. A system identification (SI) approach is used to let the algorithm get familiar with the system it is controlling. In SI, the system is excited with a known disturbance on the control commands. For large systems, it is important to construct information-rich signals that are able to persistently excite all relevant frequencies. A popular choice is a random binary signal (RBS). This is a key point of the identification process because it is possible for the algorithm to get stuck in the null space of the instantaneous prediction matrices. \edited{For example, if the system starts with a zero prediction matrix and no control signal is injected, the algorithm will think that all temporal effects are intrinsic to the turbulence. This will keep the $B$ and $C$ matrices at zero, which means that the controller will not be learned regardless of the amount of training steps.} RBSs will allow the control algorithm to explore all possible states and push the controller out of \edited{such} null spaces. Therefore, we start the control loop with an additional exploration signal on top of the atmospheric disturbance. In principle, this can also be done during the day to minimize the loss of observing time during the night. The identification is most efficient if the amplitude of the RBS signal is larger than the measurement noise.

\begin{figure}
    \begin{center}
        \includegraphics{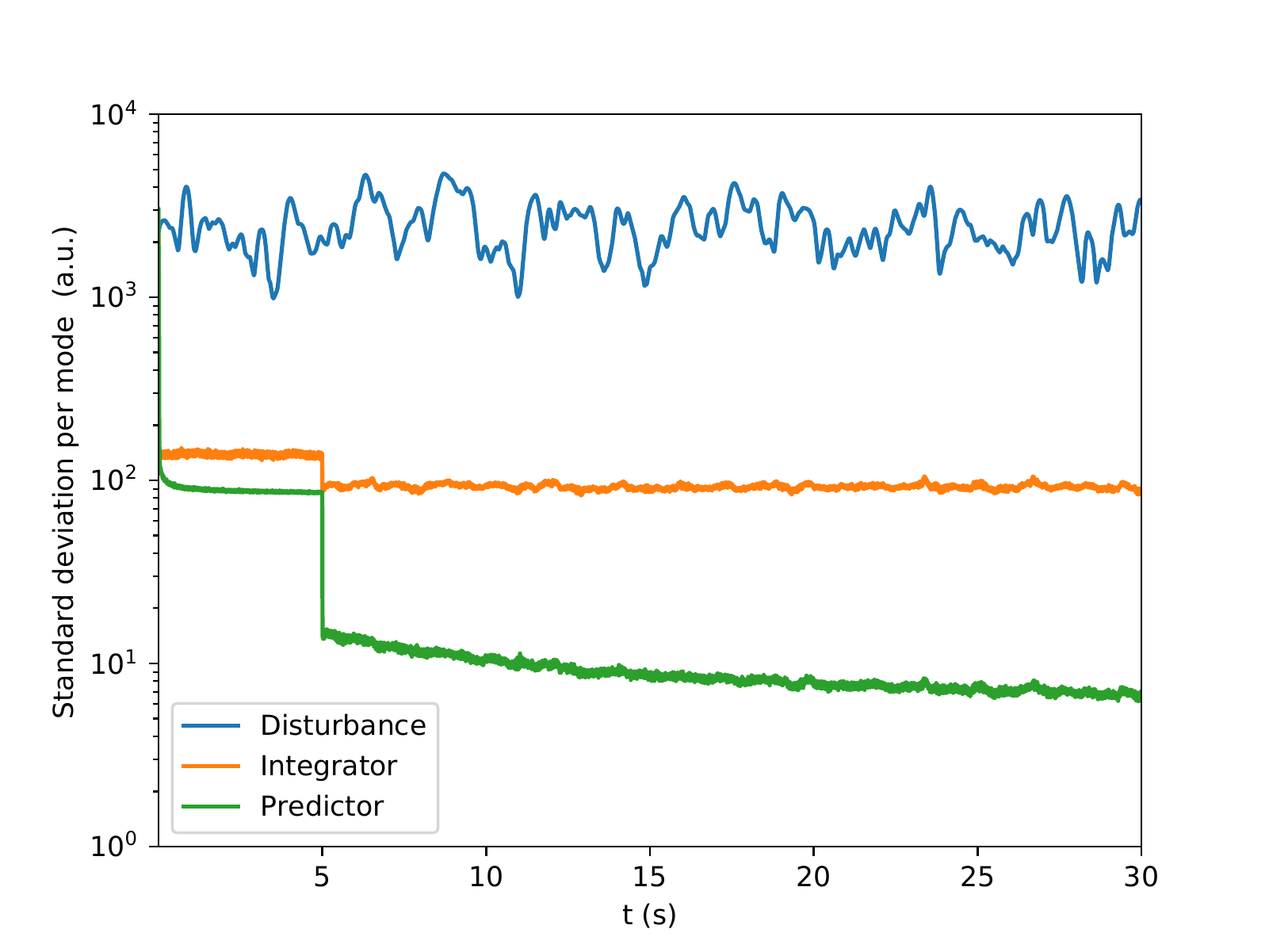}
	\end{center}
    \caption{The standard deviation of the residual modal coefficients during closed-loop operations. The first 5 seconds were used for exploration of the system with a RBS, which adds additional noise to the residuals. The influence of the RBS noise is visible through the drop of the RMS after the exploration is stopped. The DDSPC controller has learned most of the structure of the system within the first 100 ms where the RMS of the DDSPC crosses the RMS of the integrator.}
    \label{fig:learning_turbulence} 
\end{figure}

An example of the identification process can be seen in Figure \ref{fig:learning_turbulence}. The initial RMS of the DDSPC shoots up because the system is not yet known. After several tens of iteration, the controller is already performing better than an integral controller. Even with the RBS included, the DDSPC has lower RMS than a classical integrator. \edited{The integrator does not need to train, so the RBS is not necessary. The RBS signal was still injected to create an equal disturbance between both controllers.} After the exploration period has ended at the 5 second mark, we see that the predictor is performing substantially better than an integrator. It is important to mention here that the considered RMS is not the RMS of the wavefront, but of the reconstructed modal coefficients. This means that the contribution of higher-order modes \edited{, e.g. the fitting error,} has not been accounted for and the RMS in the figure can not be used to determine the improvement in Strehl. The RMS can however be used to determine the contrast improvement because the modal coefficients determine the RMS within the control radius to first order. The results show that the RMS is reduced by more than an order of magnitude, which implies an improvement of almost a factor 100 in contrast. This is the major gain that predictive control can bring for high-contrast imaging; while the Strehl improvement is not large, there can be a substantial improvement in the contrast gain. This has been shown before in previous work \cite{guyon2017eof, correia2017dkf, males2018lpc}.

\begin{figure}
    \begin{center}
        \includegraphics{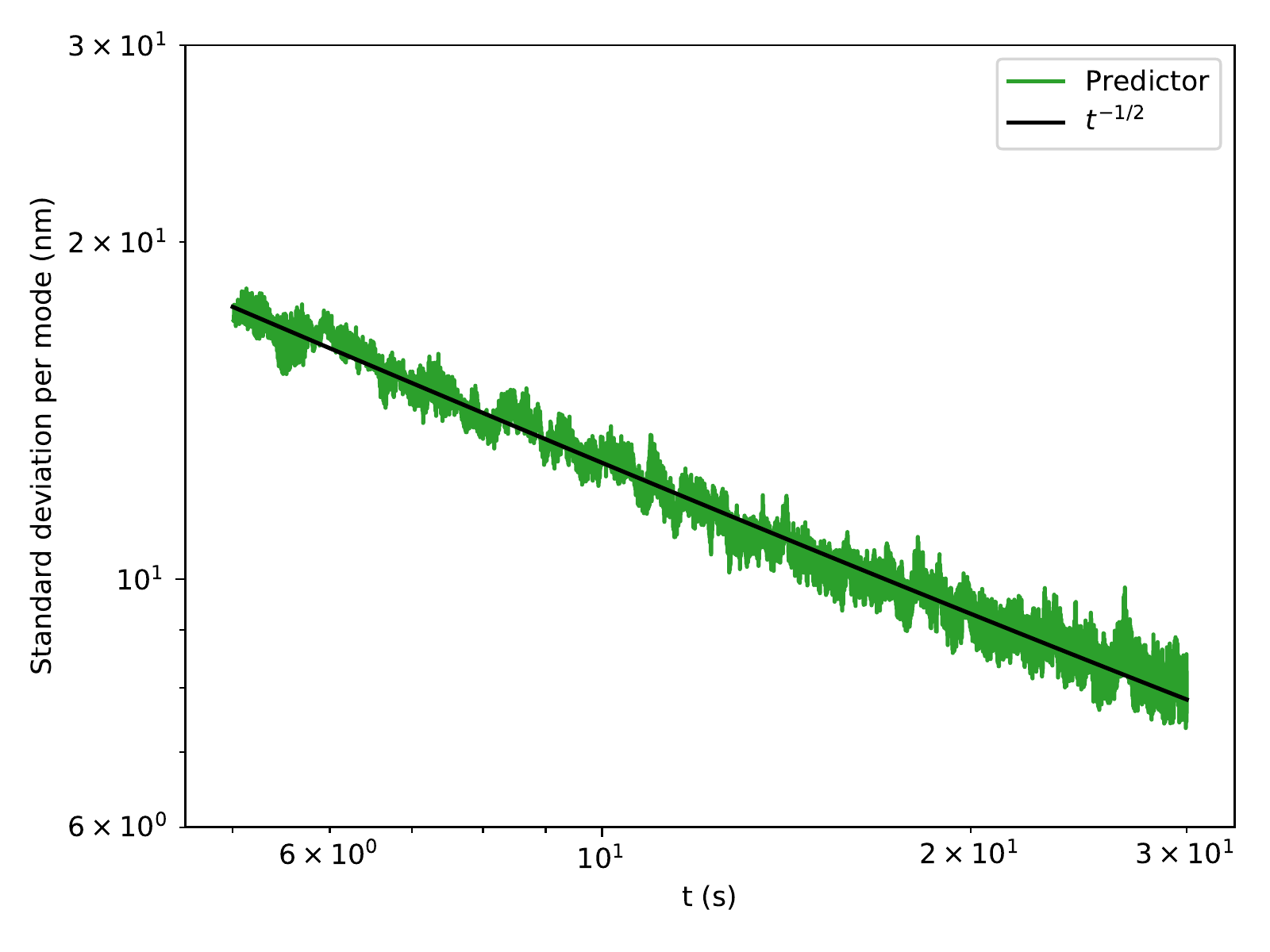}
	\end{center}
    \caption{A log-log plot of the RMS of the residual modal coefficients between 5 and 30 seconds after starting the control. In black we show the expected statistical learning curve that scales as $1/\sqrt{t}$. The RMS of the DDSPC follows the expected learning curve even after more than 29000 iterations. This indicates that the algorithm will keep learning to improve the RMS over time until all correlations are learned. }
    \label{fig:algorithmic_scaling} 
\end{figure}

After the exploration period, the DDSPC keeps learning how to improve the RMS even further, as can be seen in the part between 5 seconds and 30 seconds of Figure \ref{fig:learning_turbulence}. The region from 5 to 30 seconds are plotted separately in Figure \ref{fig:algorithmic_scaling} on a log-log scale to demonstrate the scaling behaviour. A second line that scales as $1/\sqrt{t}$ has been added for comparison. From this figure, we can see that the DDSPC algorithm is learning close to the rate of $1/\sqrt{t}$. The scaling also puts limits on how fast data-driven algorithms, such as the DDSPC, can adapt to changing conditions and the precision with which it adapts on that timescale. It is not clear from the data when DDSPC stops learning, a considerably longer time series is necessary to see if the model becomes more accurate or not and at what time it stops learning. This was not explored due to the required computation time.


\section{Adaptive optics simulations}
In this section, the performance of DDSPC is tested with end-to-end Monte Carlo simulations. We consider two cases: stationary and non-stationary frozen-flow turbulence. Both simulations use a \edited{direct wavefront sensor with infinite SNR}. All simulations make use of the High-Contrast Imaging for Python (HCIPy) package \cite{por2018hcipy}. We simulate an AO system similar to MagAO-X, see Table \ref{tab:parameters} for all parameters. The temporal response of the DM is simulated as a step response with 1 frame delay. The total system delay is then 2 frames because the act of measuring the wavefront error also delays the response by 1 frame.


\subsection{Stationary turbulence}
In this subsection, we apply the DDSPC to single-layer atmospheric disturbances which evolve according to Taylor's frozen flow\cite{taylor1938frozenflow}. The DDSPC is compared to two other controllers. The first is what we will define as the perfect controller. The perfect controller is an instantaneous controller that perfectly removes all modes from the atmosphere at each time step. The second controller is a classical integrator with a fixed gain of 0.4. For the simulations in this subsection, we will use the same parameters as Table \ref{tab:parameters} except the wind speed that will vary between simulations. For each wind speed, we simulate 5s (5000 frames) of data sampled at 1 kHz. The first 2s (2000 frames) are used for exploration with an exploration noise strength of 2 radian per mode. The history and future length of the DDSPC are 4 and 3, respectively. \edited{This choice of the horizon lengths is driven by the delay of the system. The simulated delay is 2 frames and to solve for the future command trajectory, we used a future horizon length that is 1 frame longer. Longer horizons can be used, but the states farther in the future are generally harder to predict, which will effectively introduce additional noise in the future trajectory. The history horizon was chosen based on the results from Section 2.5, which showed that longer history horizons do not necessarily improve the residuals due to over fitting. The history length was set to $N=4$ to make sure that there is more input data for the prediction than outputs that need to be predicted. The additional benefit of smaller filters is also that they require less computation time, which is sparse at operating frequencies of 1 kHz or higher. Section 2.5 also showed that the regularization parameter $\lambda$ should be set to something small, so we chose $\lambda=0.01$.}

\begin{table}[ht]
\caption{Parameters of the end-to-end adaptive optics simulations.} 
\label{tab:parameters}
\begin{center}       
\begin{tabular}{|l|l|l|} 
\hline\hline\hline\hline
\rule[-1ex]{0pt}{3.5ex}  AO Parameter & Value & Comment  \\
\hline\hline\hline\hline
\rule[-1ex]{0pt}{3.5ex}  Pupil diameter & 6.5 m &  \\
\hline
\rule[-1ex]{0pt}{3.5ex}  Actuators across pupil & 50 & Gaussian influence functions.  \\
\hline
\rule[-1ex]{0pt}{3.5ex}  Integration time & 1 ms &  \\
\hline
\rule[-1ex]{0pt}{3.5ex}  Total loop delay & 2 frames & \\
\hline\hline\hline\hline
\rule[-1ex]{0pt}{3.5ex}  Atmospheric Parameter & Value & Comment  \\
\hline\hline\hline\hline
\rule[-1ex]{0pt}{3.5ex}  r0 & 0.16 m & Median 0.62" at 0.5$\mathrm{\mu}m$ \\
\hline
\rule[-1ex]{0pt}{3.5ex}  Mean wind speed & 15 m/s &  \\
\hline
\rule[-1ex]{0pt}{3.5ex}  WFS wavelength & 800 nm &  \\
\hline\hline\hline\hline
\rule[-1ex]{0pt}{3.5ex}  Integral control parameter & Value & Comment  \\
\hline\hline\hline\hline
\rule[-1ex]{0pt}{3.5ex}  g & 0.4 & Integrator gain. \\
\hline
\rule[-1ex]{0pt}{3.5ex}  Predictive control parameter & Value & Comment  \\
\hline\hline\hline\hline
\rule[-1ex]{0pt}{3.5ex}  M & 3 & The prediction horizon. \\
\hline
\rule[-1ex]{0pt}{3.5ex}  N & 4 & The number of past samples. \\
\hline
\rule[-1ex]{0pt}{3.5ex}  $\lambda$ & 0.01 & Regularization parameter \\
\hline
\rule[-1ex]{0pt}{3.5ex}  $\gamma$ & 1.0 & Forgetting factor \\
\hline\hline\hline\hline

\end{tabular}
\end{center}
\end{table} 

\begin{figure}
    \begin{center}
        \includegraphics[width=\textwidth]{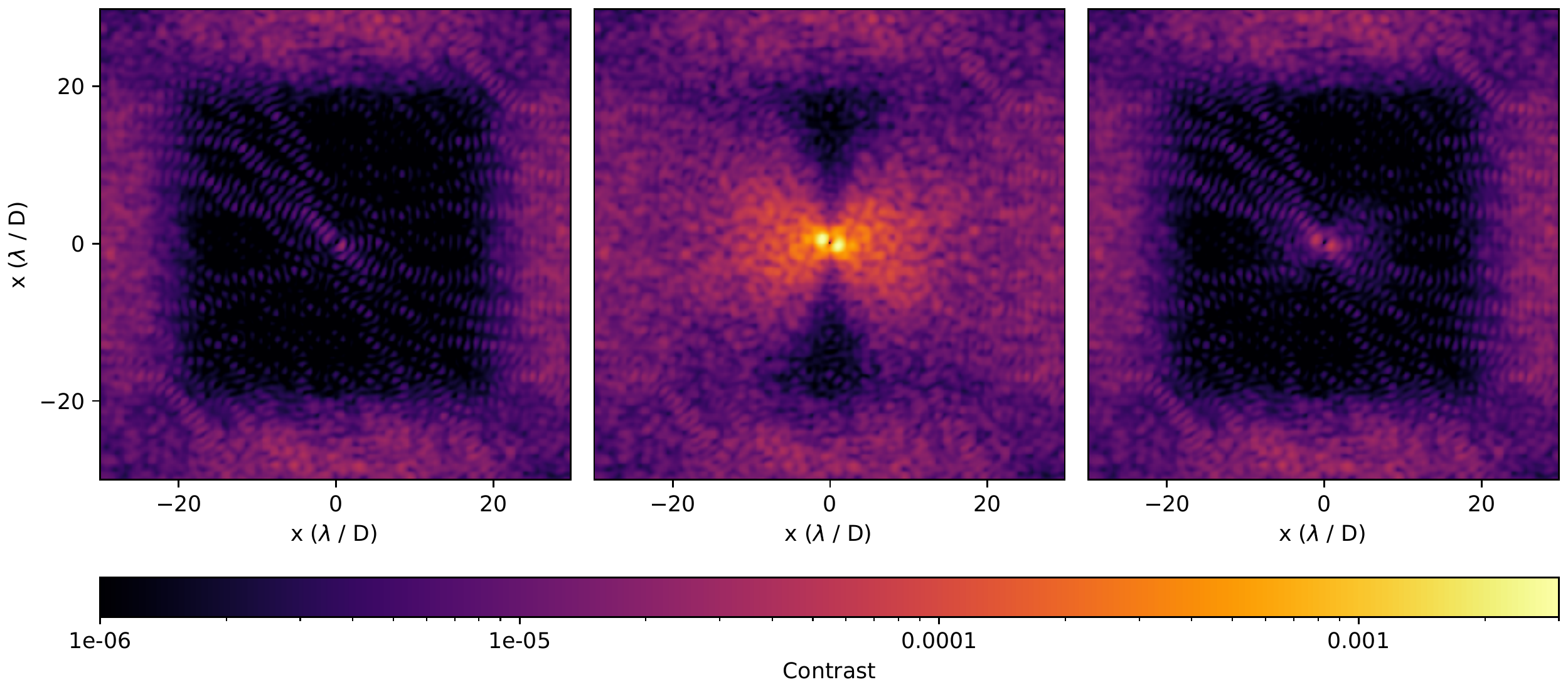}
	\end{center}
    \caption{Post-coronagraphic residuals for different controllers. The post-coronagraphic stellar residuals for a perfect controller (left). The perfect controller still shows some residuals within the dark hole that are due to spatial fitting errors. The results of a classical integrator (center) where a strong wind-driven halo is limiting the contrast. The DDSPC (right) contrast map, where the majority of the wind driven halo is removed. The contrast has improved by almost two orders of magnitude. Most of the residual structure is similar to the structure in the dark hole of the perfect controller, which indicates that the contrast is also limited by the spatial fitting errors. }
    \label{fig:focal_plane_contrast_map} 
\end{figure}

\begin{figure}
    \begin{center}
        \includegraphics[width=\textwidth]{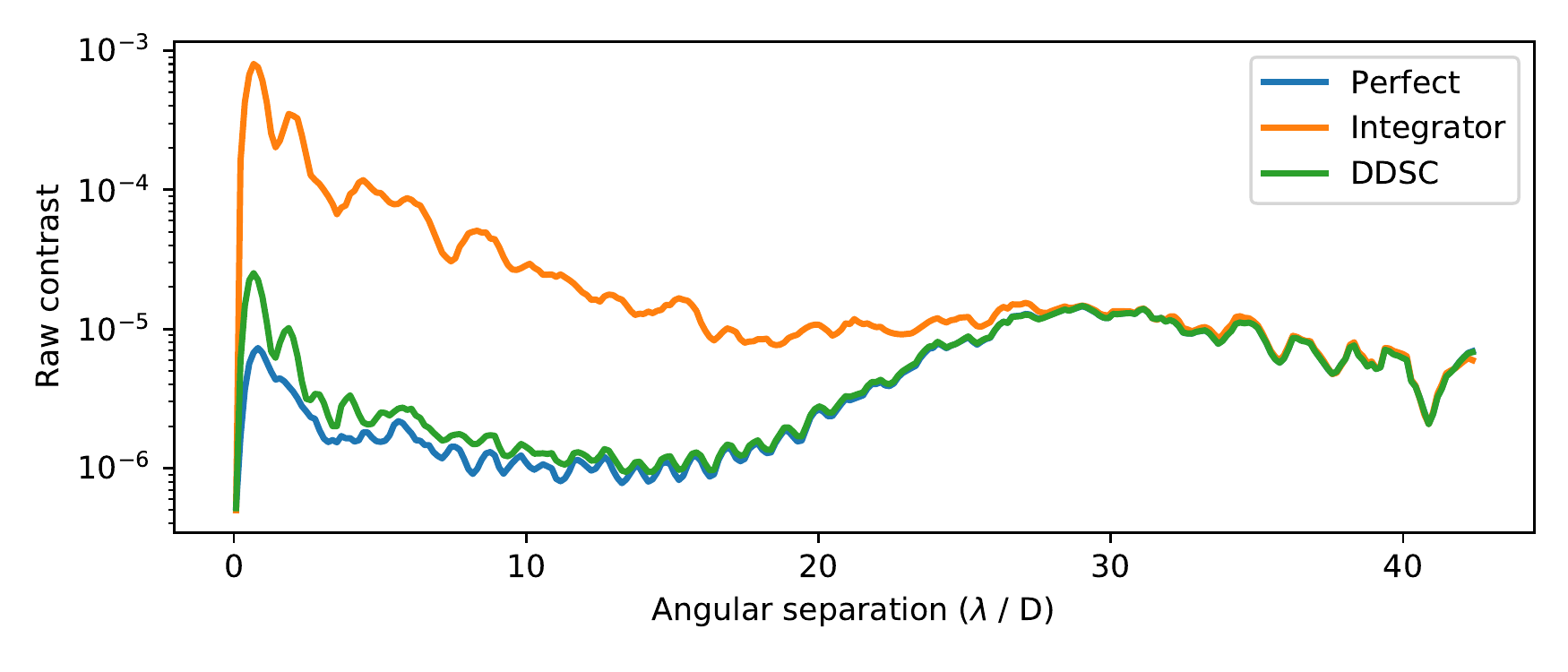}
	\end{center}
    \caption{The radial profiles of the post-coronagraphic contrast maps. The DDSPC shows two orders of improvement in contrast over the integral controller. The performance of the DDSPC at the diffraction-limit is still a factor 2 higher than the perfect controller. }
    \label{fig:radial_profile_contrast} 
\end{figure}

The wavefront residuals for each controller are propagated through a perfect coronagraph \cite{cavarroc2006coronagraph, guyon2006coronagraph} to evaluate the contrast in the focal plane. We choose to evaluate the contrast at a wavelength of 1$\upmu$m. An example of such a focal plane contrast map for a windspeed of 15 ms$^{-1}$ can be seen in Figure \ref{fig:focal_plane_contrast_map}. The contrast of the perfect controller is on the order of $10^{-6}$. There is still some residual structure in the dark-hole which comes from \edited{artifacts of the wavefront reconstruction step.} In principle, these residuals could be removed by focal-plane wavefront sensing techniques. The integrator shows the typical wind-driven halo, which in this case severely limits the contrast. The DDSPC reaches a contrast that is very similar to the perfect controller. The only difference is a small decrease in contrast at the smallest angular separations. This can be better seen in the radial profile, as shown in Figure \ref{fig:radial_profile_contrast}. The radial profile shows that the contrast of the predictive controller is within a factor of 2 of the perfect controller.

\begin{figure}
    \begin{center}
        \includegraphics{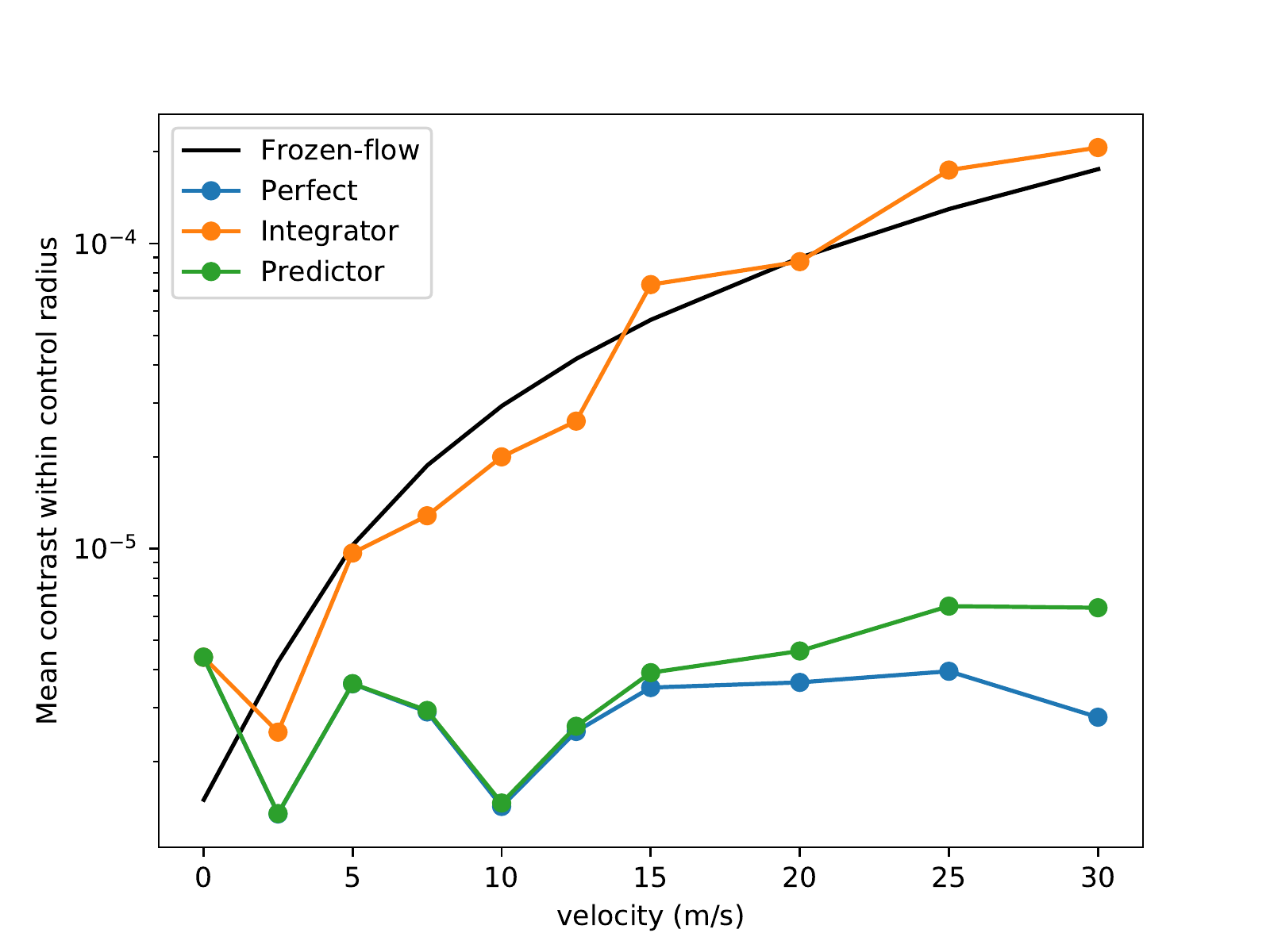}
	\end{center}
    \caption{The mean contrast within the control radius of the DM after a perfect coronagraph as function of the wind speed. The contrast is shown for the three different controllers. \edited{The black line shows the expected behaviour for an integrator as function of velocity for frozen flow turbulence based on the phase structure function. The proportionality between contrast and velocity is derived in the Section 3.1 The actual measured contrast of the integrator (orange) follows the frozen-flow relation.}The DDSPC controller stays very close to the perfect controller. Only at high wind velocities does the DDSPC start to degrade in performance, but it is still within a factor of 2 of perfect control. }
    \label{fig:frozenflow} 
\end{figure}

Even though the DDSPC does not reach the fundamental limit, it still performs order of magnitude better than the integrator. The strength of the wind-driven halo is directly proportional to the wind speed. Figure \ref{fig:frozenflow} shows how the mean contrast within the dark hole depends on the wind speed. For very low wind speeds, the integrator is reaches the same average contrast as the DDSPC. Because the atmospheric phase screen changes very slowly, the integrator has enough time to catch up. But as the velocity increases, the contrast degrades. The degradation as function of velocity follows the expected trend for Taylor's frozen flow approximation. Under Taylor's frozen flow, the phase screen is shifted by $\Delta r = \Delta v \Delta t$, while the integral controller assumes the phase screen stays static. The variance of the error is then proportional to the phase structure function because we are essentially comparing the phase screen with a shifted copy. The phase structure function itself has a power-law dependence under Taylor's frozen flow, $D_{}(\Delta r)\propto (\Delta r)^{5/3} \propto (\Delta v)^{5/3}$. The residual contrast of the integrator follows this proportionality, as can be seen in Figure \ref{fig:frozenflow}. \edited{The contrast of an optimal predictive controller will also depend on the wind speed in the same way although with a lower constant of proportionality \cite{doelman2020MNRASmintemperror}, which is why the contrast starts to degrade at higher wind speeds. At lower wind speed ($<15\mathrm{ms}^{-1}$), the DDSPC stays very close to the perfect controller.} For the DDSPC, we kept the forgetting factor at 1, which effectively means that we are keeping the full history. The performance could still be improved by optimizing the forgetting factor or adding more data (like Figure \ref{fig:learning_turbulence}).

\subsection{Non-stationary turbulence}
One of the largest challenges of the model-based predictive controllers is the effect of changes in model parameters such as wind speed and direction \cite{kooten2019wind}. Therefore, it is crucial to test the algorithm against non-stationary turbulence. Several different simulations were performed to test the robustness of the DDSPC controller.

\begin{figure}
    \begin{center}
        \includegraphics[width=\textwidth]{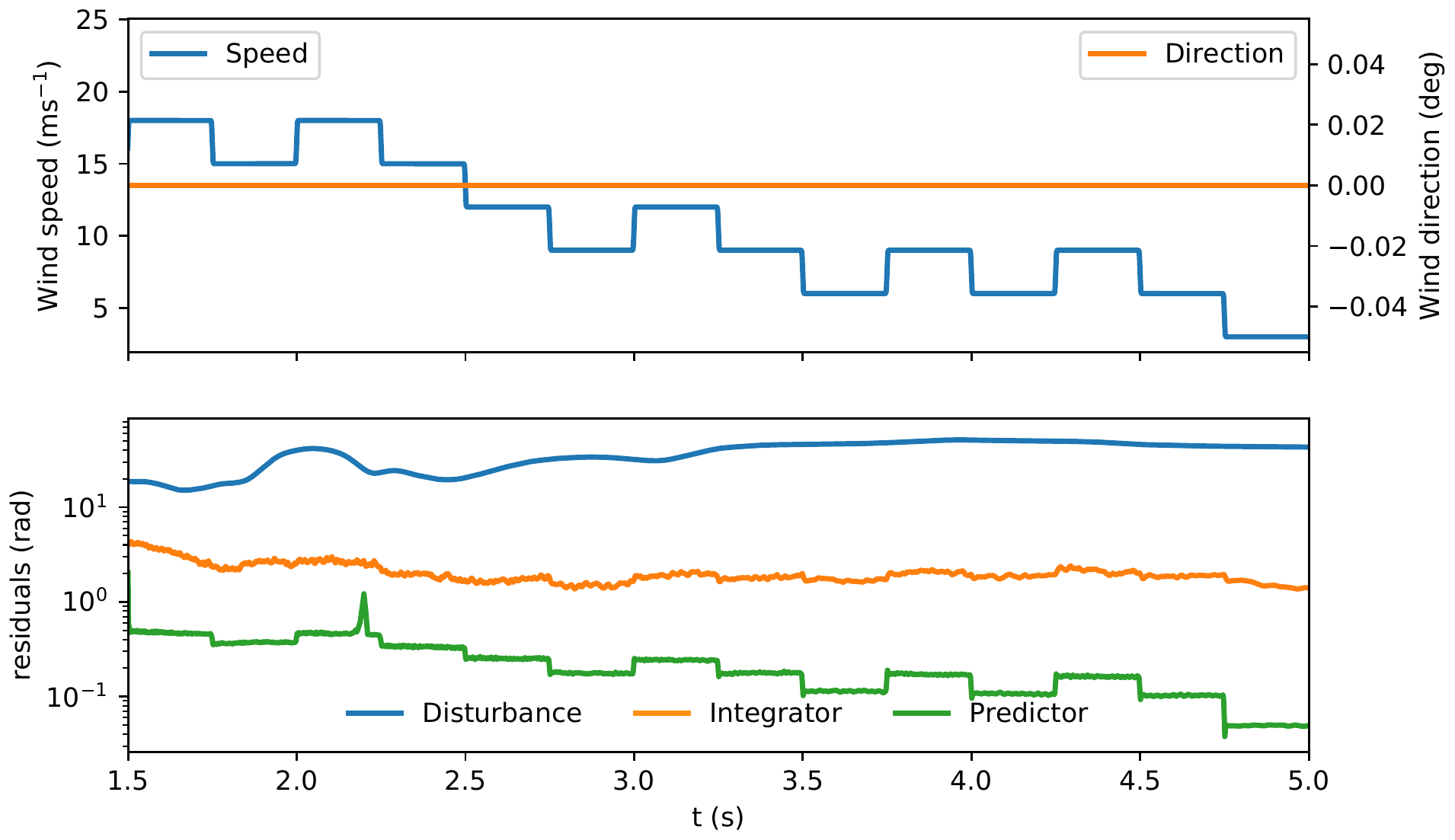}
	\end{center}
    \caption{The performance during sudden changes in wind speed. The top figure shows the wind speed and direction. The wind speed is changed by 3 ms$^{-1}$ every 250 ms. The bottom shows the RMS of the modal coefficients as function of time for the input disturbance (blue), the integrator (green) and the DDSPC controller (orange). The predictive controller has significant better performance than the integrator at all times. The performance of both the integrator and predictor is correlated with the wind speed.}
    \label{fig:non_stationary_wind_performance} 
\end{figure}

In the first simulation, the wind speed is changed by 3 ms$^{-1}$ every 250 ms. Whether the speed increases or decreases is randomly chosen. Except for the wind speed, all other parameters of the simulation are still the same as Table \ref{tab:parameters}. The results can be seen in Figure \ref{fig:non_stationary_wind_performance}. The first 1.5 seconds are used for training the controller and are not shown. The DDSPC controller is systematically performing better than the integral controller by 1 or 2 orders of magnitude. The RMS of the DDSPC is strongly correlated with the wind speed, which is expected based on the structure function. The integrator RMS is also correlated with the wind speed, but the relative changes are smaller than for the DDSPC controller. This indicates that a large fraction of the RMS of the integrator consists of the temporal dynamics of the DM. \edited{For the simulations here, the temporal dynamics of the DM is the delay because no additional temporal dynamics are included. This shows that the non-stationary wind velocities are not the limiting factor, but the delay itself.}

\begin{figure}
    \begin{center}
        \includegraphics[width=\textwidth]{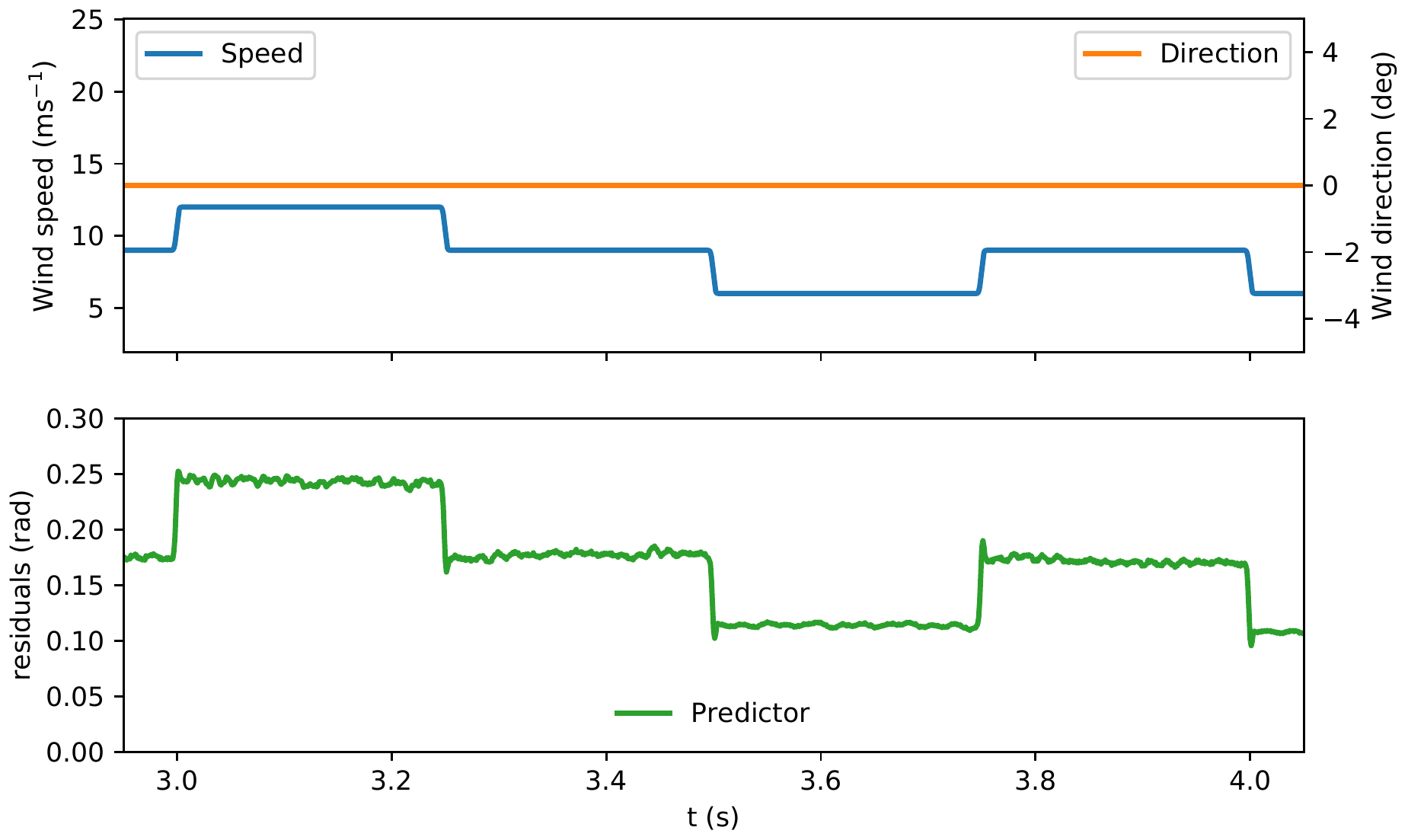}
	\end{center}
    \caption{A zoom in on the predictive controller residuals during step changes in the wind speed. The top figure shows the wind speed and direction. The bottom figure shows the RMS of the modal coefficients. There is a strong correlation between the wind speed and the controller performance. The response during a step change shows a small over or undershoot which settles within 3 frames. The step response shows behaviour similar to the Gibbs phenomenon.}
    \label{fig:wind_speed_step_response} 
\end{figure}

Figure \ref{fig:wind_speed_step_response} shows a zoom-in of the response of the DDSPC between 3s and 4s. This highlights the wind step response of the controller. We see that slight over and undershoots happen when the wind speed suddenly changes. The residuals quickly converge within 4 frames, which is the length of the history that is being used. When the wind speed changes, the history vector contains information about the wind speed before and after the jump. Only after 4 frames has the previous wind speed been completely removed from the history vector. Therefore we see that the RMS converges in about 4 frames.

The over and undershoots can also be explained from the frequency domain. The DDSPC controller uses the knowledge of the system dynamics to flatten the PSD of the disturbance. An optimal predictive controller \edited{tries to bring the output PSD as close as possible to a white output spectrum}. This means that a predictive controller acts similar to a temporal deconvolution. Because we only approximate the system by a finite \edited{number of samples (the history length)}, we are susceptible to the Gibbs phenomenon. The Gibbs phenomenon shows up when a discontinuity is encountered and creates strong oscillations at the discontinuity. These initial simulations show that the DDSPC can handle sudden changes in the atmosphere quite well and only very few frames are required to take the new conditions into account.

\begin{figure}
    \begin{center}
        \includegraphics[width=\textwidth]{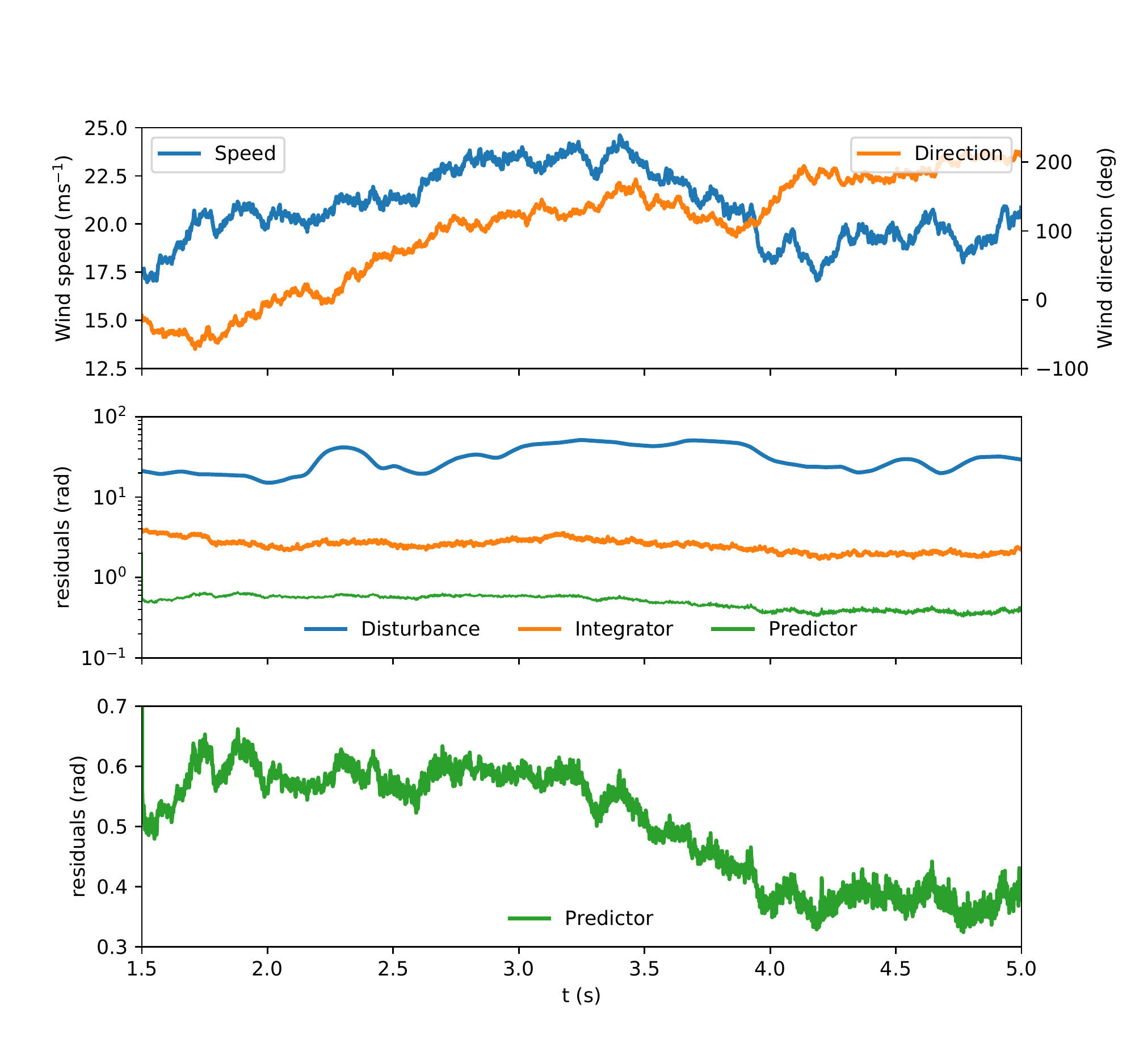}
	\end{center}
    \caption{The performance with a non-stationary disturbance. The top figure shows the wind speed and direction, which are both randomly changed at each time step. The bottom shows the RMS of the modal coefficients as function of time for the input disturbance (blue), the integrator (green) and the DDSPC (orange). The predictive controller has significant better performance than the integrator at all times. The performance of both the integrator and predictor is correlated with the wind speed.}
    \label{fig:white_noise_wind_response} 
\end{figure}

The second simulation we perform includes both changes in the wind speed and wind direction. To stress-test the controller, we decided to create completely random changes in the wind speed and direction. Both variables perform a random walk. The average wind speed is 21.0 ms$^{-1}$ with a standard deviation of 3 ms$^{-1}$. The wind direction changes by more than 270 degrees during the simulation. This is a significant amount over 5 seconds. Because there is no pattern to the wind speed changes, \edited{the changes in the atmospheric disturbance become more random. This make it more difficult to predict. Consequently, the predictive controller will show less gain over the integrator.} The results are shown in Figure \ref{fig:white_noise_wind_response}. The predictive controller still performs better than the integrator, but the difference is smaller than in the previous test. We do however still see improvement as the time increases. The wind speed is relatively constant at 20 ms$^{-1}$ around the 2s mark with a residual RMS of 0.6 radian. At 4.5s the wind speed is back to 20 ms$^{-1}$ again but this time the residuals are around 0.35 radian RMS. \edited{ The improvement is roughly a factor of 1.7. This improvement is very similar to the expected gain based on the learning rate, $\sqrt{t}$, from Section \ref{sec:learning}. That learning rate predicts a gain of $\sqrt{\frac{4.5}{2.0}}\approx1.5$.} The difference can be explained by the small differences in the actual wind speed. The wind speed is a little bit higher on average at 2s if we compared it against the wind speeds at 4.5s. \edited{This suggests that the controller has been constantly learning over this period, even under varying conditions.} Globally, the residuals show that the performance is correlated with the wind speed similar to the behaviour in the wind speed step response simulation.

We have not explicitly modelled the effects of changing $r_0$, but this was included in all simulations. Due to the way the phase screens are generated, the $r_0$ that is specified is a temporal average, which means that there are variations in time. Therefore, we assume that explicitly adding non-stationarity of $r_0$ will not influence the presented results. The simulations show that the DDSPC controller can handle non-stationary atmospheric turbulence. The second set of simulations shows that even completely random changes in the conditions can be handled. The actual wind speed has more temporal structure \cite{kooten2019wind}, therefore we expect that better performance can be achieved with more realistic wind models.

\section{Low-order control verification with MagAO-X}
\subsection{Setup description}
\begin{figure}
    \begin{center}
        \includegraphics[width=\textwidth]{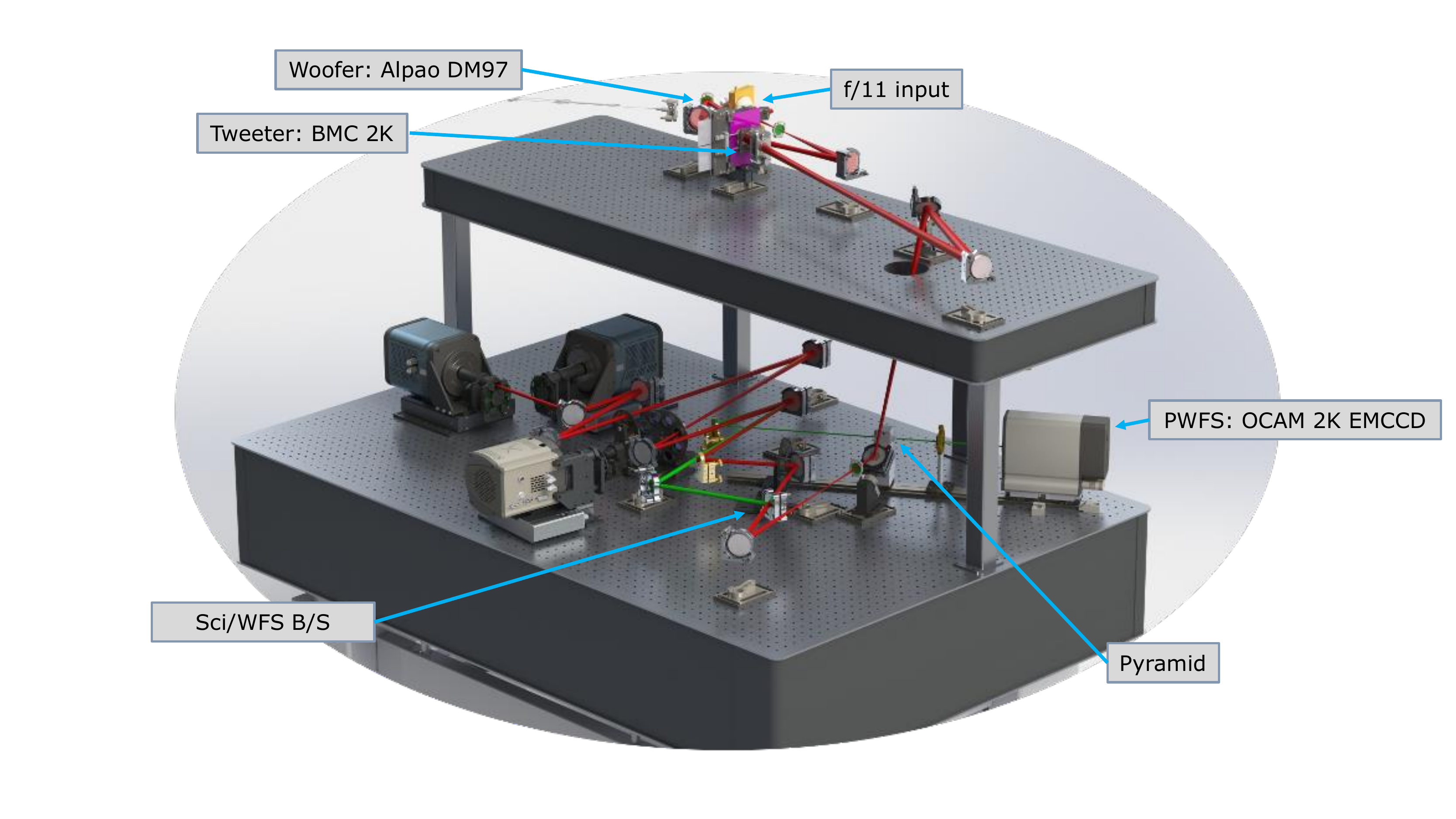}
	\end{center}
    \caption{A solid works render of MagAO-X. On the top bench, the beam enters at f/11 and then passes the woofer DM and the tweeter DM. After the periscope, the beam is split into a science path and a wavefront sensor path (green rays). The green rays hit the pyramid prism and the subsequent pupils are imaged onto the OCAM 2K. Next to the pyramid prism is the CAMTIP camera, which is not shown in the figure, that monitors the PSF that hits the tip of the pyramid prism.}
    \label{fig:magaox} 
\end{figure}

In this section, we verify the DDSPC algorithm in the lab with MagAO-X. A schematic of MagAO-X can be seen in Figure \ref{fig:magaox}. MagAO-X uses a woofer-tweeter architecture with an ALPAO-97 DM as woofer and a Boston Micromachines 2K tweeter\cite{males2018magaox, close2018magaox}. The system accepts an f/11 beam that first hits the woofer and then the tweeter. This beam is relayed to the lower bench where the pyramid wavefront sensor (PWFS) and the science cameras sit. \edited{ Similarly to previous tests of predictive control, we used the woofer DM as the disturbance creator by running simulated phase screens across the DM \cite{jensenclem2019keck}. In all our tests, we use the PWFS without modulation and with a disturbance outside of the linear regime to show that non-linearity is not a problem for the proposed algorithm. The main advantage of the DDSPC algorithm is that it operates in closed-loop. In closed-loop the residual wavefront errors are small and within the linear range of the PWFS.} We monitor the PSF with the CAMTIP camera, which monitors the (modulated) PSF that hits the pyramid prism. All tests were done with an i' filter that transmits the wavelength range from 700 nm to 820 nm. The internal light source of MagAO-X creates a brightness equivalent to a 0th magnitude star in the i' filter.

The current implementation of the control algorithm is written in Python, which limited the number of actuators that we could control and the loop speed at which we could run. Therefore, we tested the algorithm at a frame speed of 200 Hz with the 97 actuator woofer DM. To calibrate the response of wavefront to the woofer's actuators, we applied 2000 random patterns on the DM with a surface RMS of 40 nm, which becomes 80 nm in total reflected wavefront RMS. This data set was then divided into a training part (1500 samples) and a validation part (500 samples). From the training data we derived the \edited{reconstruction} matrix as,
\begin{equation}
    M = V_TS_T^T(S_TS_T^T + \alpha I)^{-1}.
\end{equation}
With $M$ the \edited{reconstruction} matrix, $V_T$ the applied commands of the training data, $S_T$ the measured slopes of the training data, and $\alpha$ the regularization parameter. The optimal regularization parameter is chosen as the $\alpha$ which minimizes the residuals of the reconstructed coefficients in the validation dataset,
\begin{equation}
    \argmin_{\alpha} |V_V - M\left(\alpha\right) S_V|^2.
\end{equation}
Here, $V_V$ and $S_V$ are the applied commands and measured slopes of the validation data set. After finding the optimal regularization parameter, we removed the actuators that were not fully illuminated, which reduced the number of actuators that were controlled from 97 to 86. \edited{For the illuminated actuators, the reconstruction residuals were below 10\,\%}.

The predictive controllers are initially trained by applying a RBS of 2000 samples in time, with an individual random series for each actuator. Because the controller does not know anything about the system, we started the controller with a strong regularization of $\lambda=10$. This heavily penalizes the system for any control command, which is desirable if nothing is known yet. After the first round of RBS, we lowered the regularization parameter by a factor 100 and ran another RBS. And finally, we reduced $\lambda$ again by another factor of 100 and did the last training iteration. The final value of $\lambda=0.001$ was used for the rest of the tests.

\subsection{Results}
\begin{figure}
    \begin{center}
        \includegraphics[width=\textwidth]{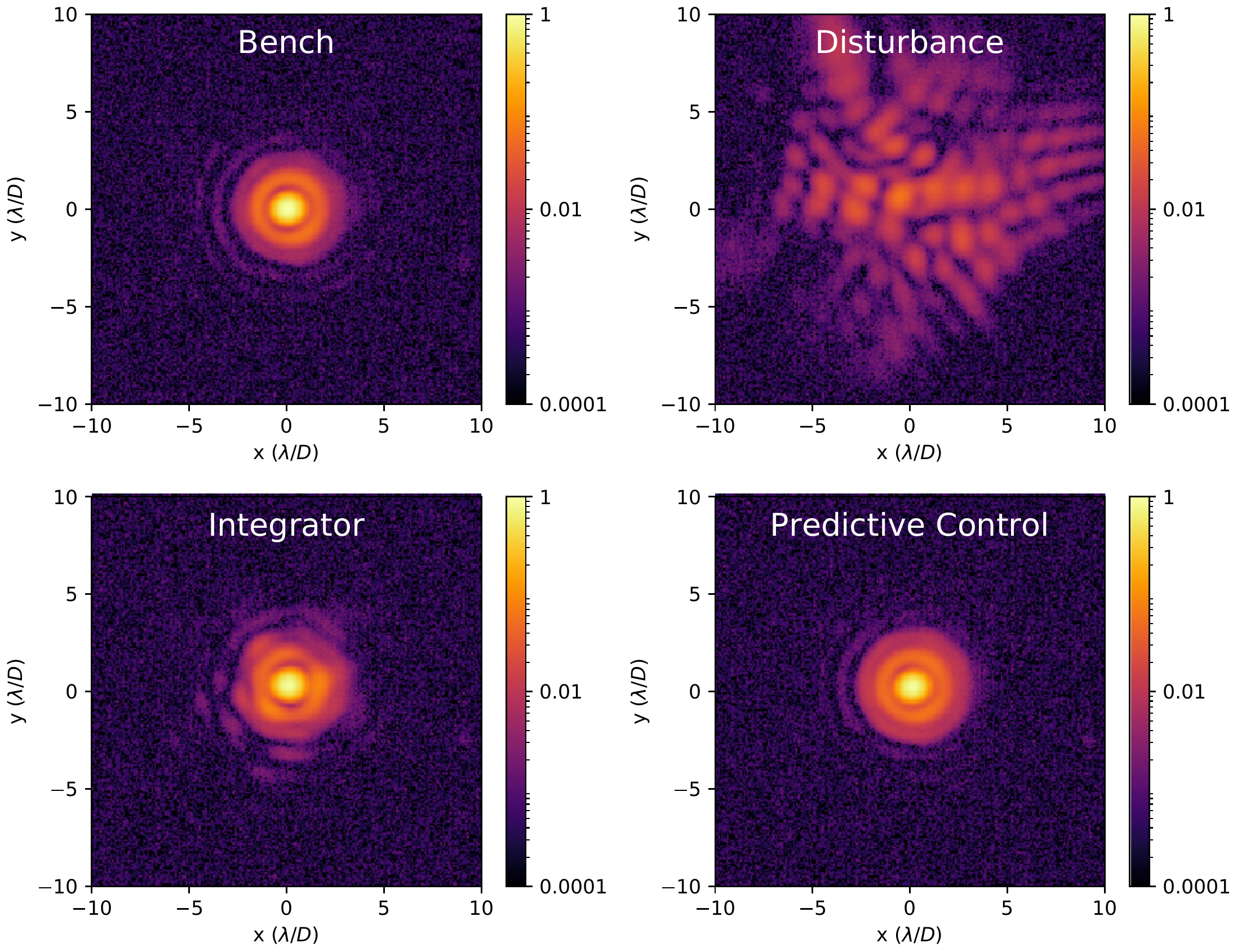}
	\end{center}
    \caption{The PSF during the lab experiments \edited{when runnning a 15 ms$^{-1}$ wind across the DM}. The top left figure shows the PSF of the system at rest, without any disturbance or control. There is a visible defocus of the PSF. The top right is a snapshot during the replay of the generated atmospheric phase screens. The bottom left shows the instantaneous PSF with an integral controller with a gain of 0.3. The integrator PSF still has distortions of the Airy rings. The bottom right shows the predictive control PSF which has no visible distortions. A video of the PSFs is available online (MP4, 11 Mb).}
    \label{fig:snapshot} 
\end{figure}

A series of Kolmogorov phase screens generated with HCIPy \cite{por2018hcipy} that evolved according to frozen flow were used to mimic atmospheric disturbances. The phase screens were created with a $r_0=0.16$m and $L_0=40$m. The PSF was recorded by the CAMTIP camera at 30 Hz while the phase screens ran across the DM. Examples of the instantaneous PSFs \edited{for a wind speed of 15 ms$^{-1}$}can be seen in Figure \ref{fig:snapshot}. There is a defocus visible in the PSFs because the CAMTIP camera is slightly out of focus. The \edited{initial} PSF without any correction is heavily distorted and well outside of the PWFS linear regime. The non-linearity is an important aspect to test for predictive control because previous data-driven methods relied on the open-loop reconstruction of the wavefront \cite{guyon2017eof, jensenclem2019keck} and the non-linearities will introduce strong model errors \edited{due to saturation or optical gain variations.} Therefore, it is important to test whether the proposed DDSPC will be influenced or not by non-linearity. Looking at the PSF of the integrator shows that there are residual aberrations, while the PSF of the DDSPC looks perfect by eye. This first image already demonstrates the performance boost of DDSPC, even with the non-linearity of the PWFS included. 

\begin{figure}
    \begin{center}
        \includegraphics[width=\textwidth]{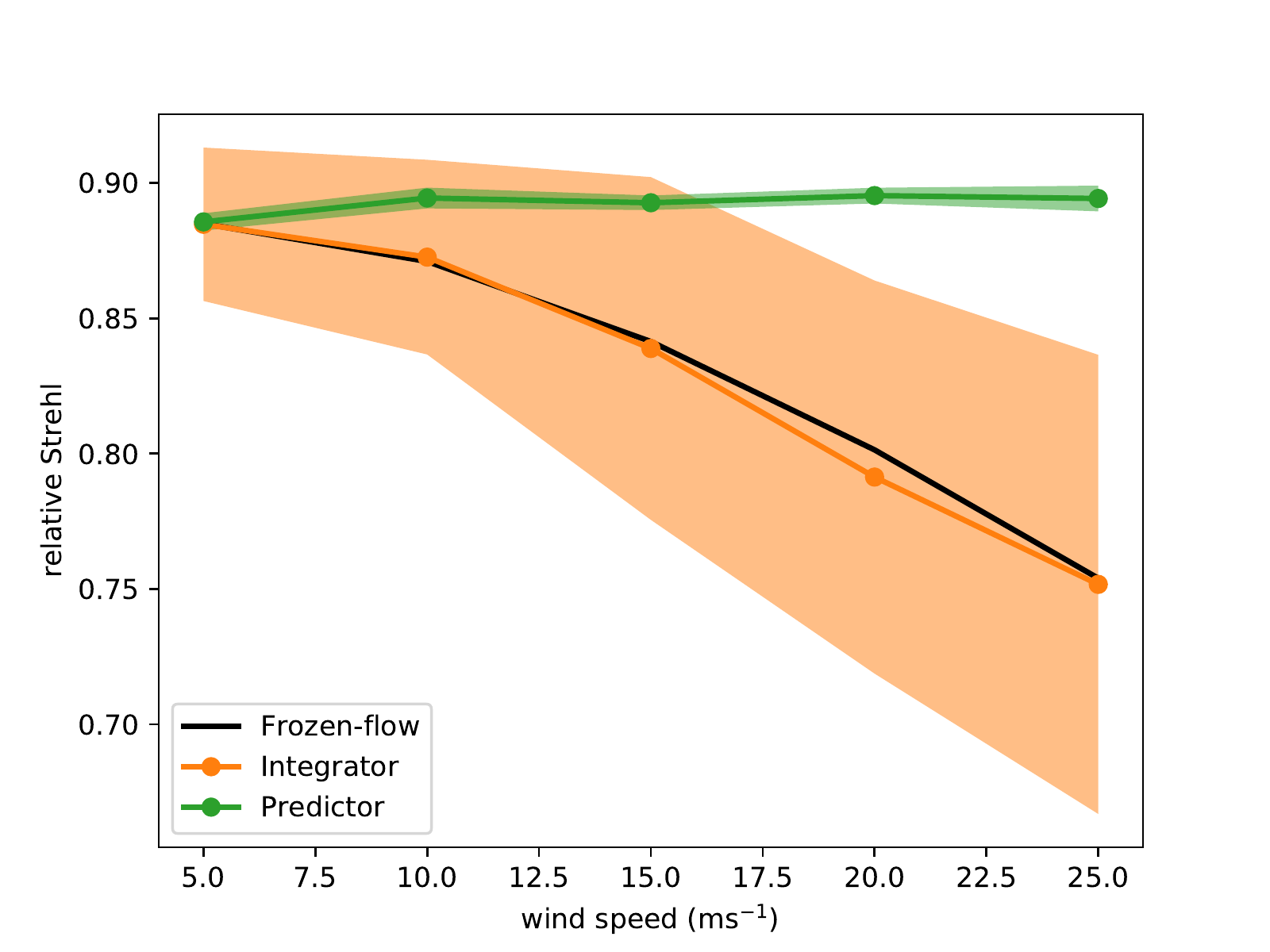}
	\end{center}
    \caption{Strehl of the measured PSFs relative to the bench PSF versus wind speed. The shaded area around each curve is the 16th and 84th percentile of the Strehl in time and corresponds to the temporal stability of the PSF. The median performance of the integrator follows the theoretical Strehl according to frozen-flow. The Strehl of the predictive controller has a smaller spread than the integrator and is independent of wind speed.}
    \label{fig:strehl_timeseries} 
\end{figure}

Figure \ref{fig:strehl_timeseries} demonstrates the performance of both the integrator and DDSPC under a range of wind speeds. The integrator's declining performance with increased wind speed follows the expectation from frozen flow. On the other hand, the performance of the DDSPC seems to be independent of the wind speed. Another major benefit is also apparent in the stability of the Strehl, the DDSPC has a significant smaller spread than the integrator. This spread directly translates into PSF stability, and a more stable PSF will lead to an increased sensitivity to faint companions after PSF subtraction techniques. However, Figure \ref{fig:strehl_timeseries} also shows that the Strehl of the integrator is sometimes higher than that of the predictive controller. We think that this is due to the focus offset of the CAMTIP camera. The DDSPC creates a well-corrected wavefront on the PWFS and due to the focus NCPA, the CAMTIP PSF will not have the highest possible Strehl. The integrator is not able to reach this performance of control and will have some left over defocus that can compensate for the focus NCPA. Therefore, this creates a situation where a less-performing controller can reach a higher Strehl.

\begin{figure}
    \begin{center}
        \includegraphics[width=\textwidth]{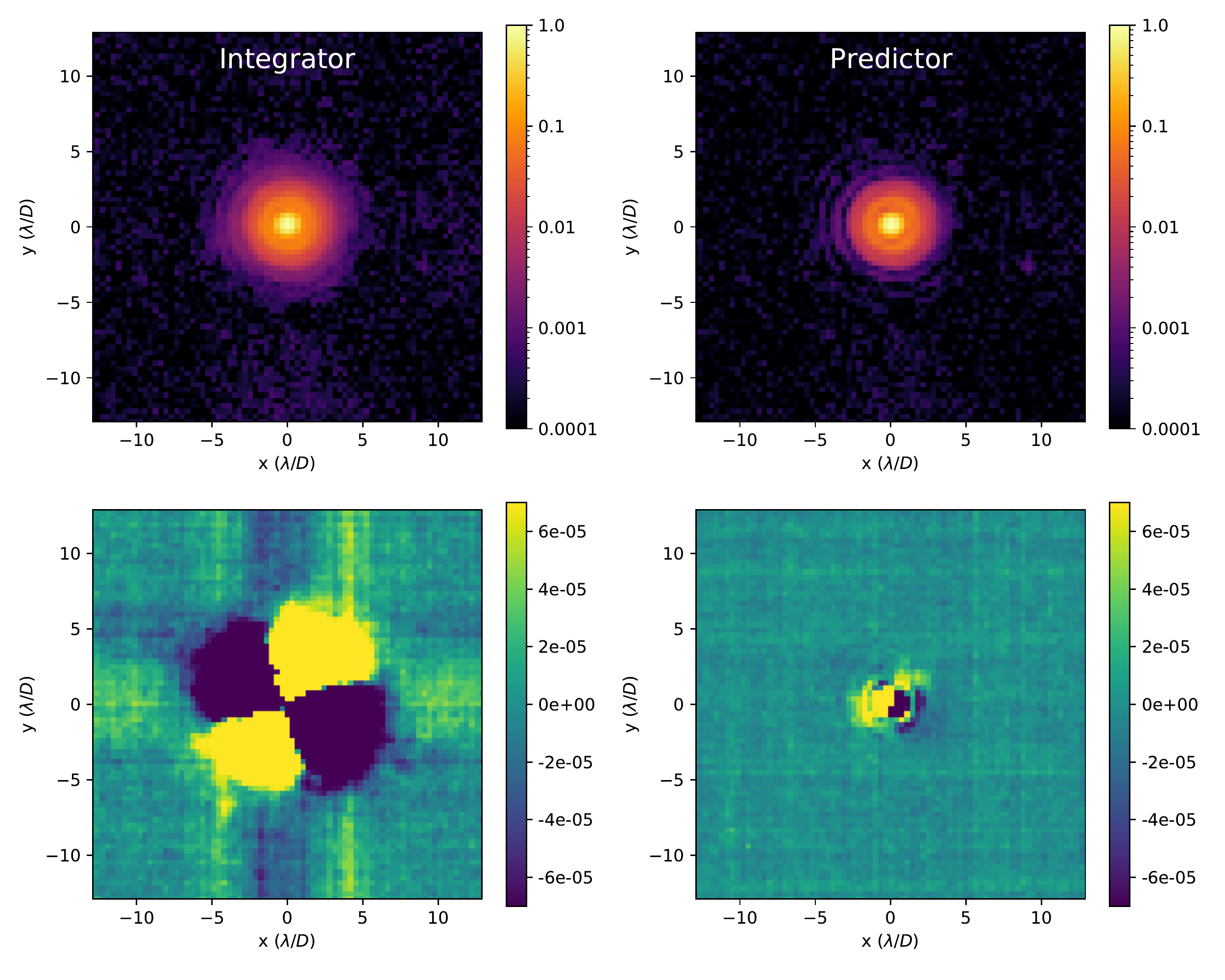}
	\end{center}
    \caption{The measured PSFs (top) and post-processed residuals (bottom) for a wind speed of 25 ms$^{-1}$. The left column shows the results for the integrator while the right shows the results for the predictor. \edited{The control region of the DM extends to 5.5 $\lambda$.} \edited{The integrator shows a strong wind-driven halo, while the halo is gone with the DDSPC.}}
    \label{fig:focal_plane_contrast} 
\end{figure}

A better performance metric would be the post-processed contrast that can be reached. \edited{At the time of the experiments, MagAO-X did not have coronagraphs installed, so it was not possible to directly measure the raw contrast.} To \edited{still} estimate the gain in contrast between the two controllers, we created two data sets for each wind speed. The first data set has a wind moving horizontally, while the second has a wind moving vertically. \edited{The temporal error of the integrator will create a wind-driven halo that is oriented in the direction of the wind (e.g. see Figure \ref{fig:focal_plane_contrast_map}). Because the wind is rotated between the exposures, the wind driven halo also rotates. This creates a typical butterfly pattern that decrease the post-processed contrast \cite{cantalloube2020windhalo}. The difference between the two data sets will remove the static diffraction pattern and show the residuals due to the time lag.} An example of the post-processed focal plane is shown in Figure \ref{fig:focal_plane_contrast} where we show the residuals for a wind speed of 25 ms$^{-1}$. For both wind directions, we stack all PSFs in time (10 s of total integration time) to create a single PSF per wind direction. These two PSFs are then combined to estimate the average PSF. The diffraction pattern is removed by taking the difference between the PSFs of the two wind directions. 


\begin{figure}
    \begin{center}
        \includegraphics[width=\textwidth]{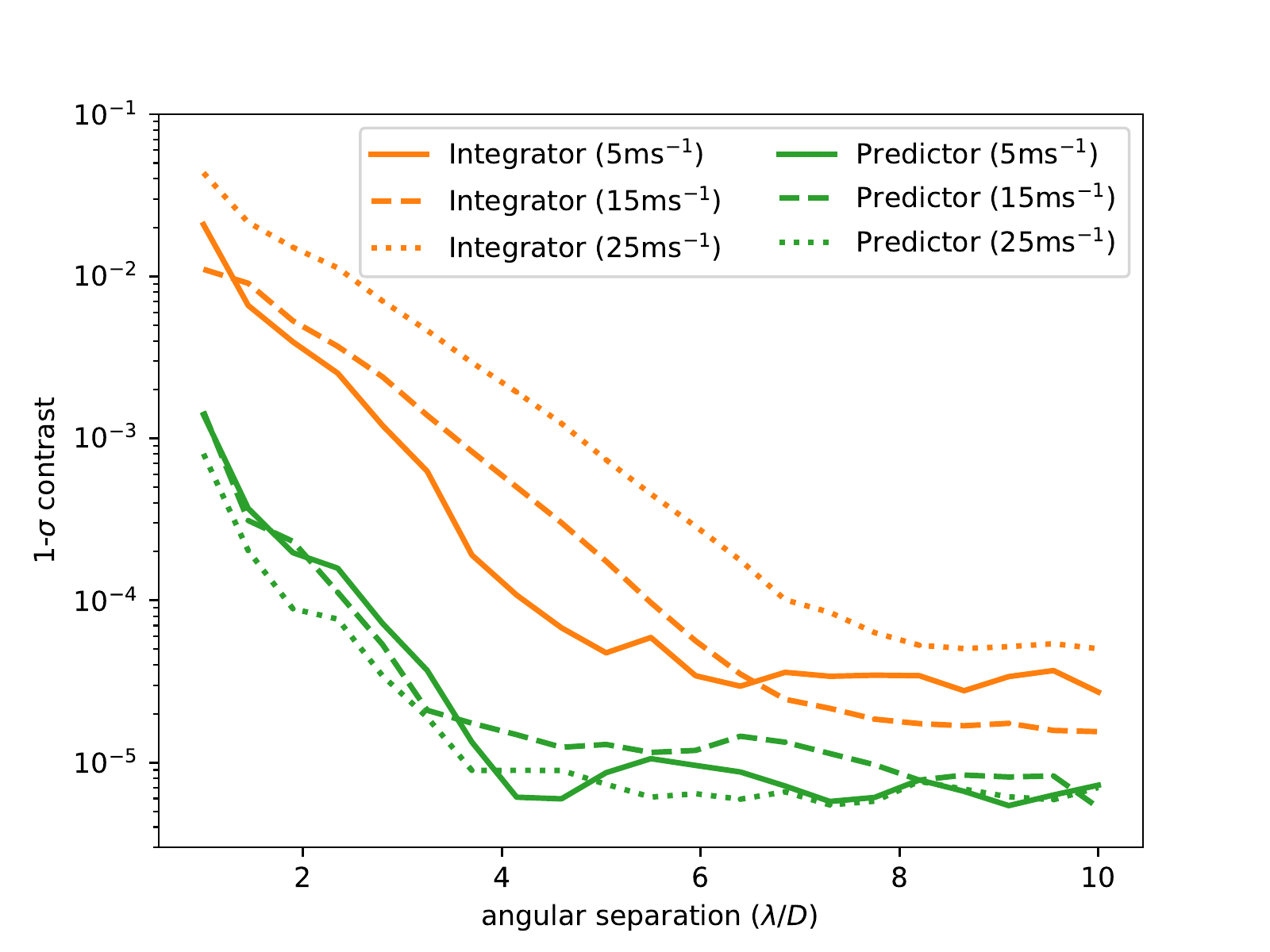}
	\end{center}
    \caption{The contrast curve for different wind speeds. The integrator is shown in orange, while the predictor is shown in green. The different line styles correspond to different wind speeds. \edited{The contrast curve for both controllers flattens out at the larger angular separations due to detector noise}. In general, we see a much deeper contrast for the DDSPC than for the integrator. The achieved contrast of DDSPC is almost independent of the wind speed, similar to the behavior of the Strehl. For the integrator, we see a clear degradation of the contrast with increasing wind speed.}
    \label{fig:lab_contrast_curves} 
\end{figure}

To quantify the contrast gain we determined the post-processed contrast curve for several wind speeds with aperture photometry. An aperture diameter of 1.4 $\lambda/D$ was used. The contrast curves are shown in Figure \ref{fig:lab_contrast_curves} for three different wind speeds, 5 ms$^{-1}$, 15 ms$^{-1}$ and 25 ms$^{-1}$. The mean wind speed at the Las Campanas Observatory site is 18 ms$^{-1}$ \cite{males2018magaox}, which means that the three chosen wind speeds are a good range for the actual on-sky conditions. The contrast of the integrator degrades as the wind speed increases, which is similar to the simulated behaviour. The DDSPC however has close to no dependence on the wind speed. In all cases, the DDSPC has superior performance. For the low speed (5 ms$^{-1}$), the predictor gains a factor of 10 to 20 in contrast. However, for the higher speeds there is a gain of more than a factor of 100, which was the prediction for predictive control in earlier work\cite{guyon2017eof, males2018lpc, correia2020hcipwfs}. These results also show a much higher gain in contrast than earlier tests that only showed a moderate improvement of a factor of 2 \cite{jensenclem2019keck}. Our lab results clearly show the benefit of predictive control for high-contrast imaging.

\section{Discussion and conclusion}
In this paper, we presented the data-driven subspace predictive controller for adaptive optics for high-contrast imaging. The method is completely data-driven; no knowledge of the system is necessary a-priori and it learns while operating in closed-loop. To lower the computational burden, we used a distributed implementation of DDSPC where we control each mode independently. \edited{The hyperparameters of the controller have been explored for various power-law disturbances. This showed that small models result in the lowest closed-loop residuals and that regularization of the controller is only necessary when shallow power laws (or noise) are the dominant input. The stability of the controller has been verified numerically for a wide range of power laws and hyper parameters. This gives us confidence that the algorithm is stable in most situations.}

\edited{The controller was tested both in an end-to-end adaptive optics simulation with stationary and non-stationary turbulence and shows an improvement of more than 2 order of magnitude in contrast.} To verify that the algorithm can work in real systems, we used it to control the woofer of MagAO-X with a PWFS. On the bench, we have achieved a contrast improvement of a factor 10 to 20 for low wind speeds, while we gained more than two orders of magnitude for high wind speeds. This was done in combination with an unmodulated PWFS, showing that the DDSPC can work even when strong non-linearities are involved.

Several steps need to be taken to move from the bench tests to the on-sky with the DDSPC algorithm on MagAO-X. The most important step is to completely port the Python implementation to a lower-level programming language. Because the algorithm mainly consists of matrix-vector multiplications and matrix inversions, a major speed up is expected if we run the algorithm on the GPU. The first speed and timing tests with a GPU implementation of the DDSPC have already shown that it is possible to control 1600 modes faster than 1.5 kHz in double precision and faster than 3 kHz in single precision. Because of the distributed nature of the algorithm, we expect that it is trivial to switch from a single GPU to multiple GPUs, which would allow us to run DDSPC for the ExAO systems of the future GSMTs on current hardware.

The major focus in this work has been on the data-driven control to suppress the servo-lag error. The current controller does not \edited{explicitly} take into account that the measurements contain noise \edited{, however it can a handle noisy measurements because all our lab measurements contained noise. The exact behaviour of the algorithm with noisy measurements has not been looked at.} The proposed cost function minimizes the future predicted residuals, which would be the correct cost function for deterministic systems. For a stochastic system \edited{with unequal variance on the future predictions (Equation \ref{eq:cost})}, this \edited{does not create the optimal controller}. \edited{For the optimal estimator each measurement will need to be weighted by its inverse variance}, leading to a modification of Equation \ref{eq:controller}. It has been shown that the DDSPC can converge to the optimal Linear-quadratic-Gaussian (LQG) controller \edited{for an infinite prediction horizon} \cite{favoreel1999mflqg}. For this, two steps are necessary: the first step estimates the prediction matrices, while the second step applies a filter based on the singular values of the prediction matrix. This second step creates the optimal Kalman filter for the future states. Our implementation with the RLS does not apply this second filter step \edited{and retains all temporal modes}. Adding such a step may increase the robustness of the DDSPC controller \edited{against} measurement noise. Finally, we could add spatial coupling back into the control matrices. We originally chose to remove those for computational reasons, but the addition of some spatial coupling may help to reduce the effects of noise. All three modifications will lead to increased computational demands, so for future work we will need to balance the complexity of our model against available computational resources.

An on-sky demonstration of the DDSPC with MagAO-X is planned in the near future, which will be a step towards detecting Earth-like planets around nearby stars.


\subsection*{Disclosures}
The authors declare that they have no conflict of interest.

\acknowledgments 
Support for this work was provided by NASA through the NASA Hubble Fellowship grant \#HST-HF2-51436.001-A awarded by the Space Telescope Science Institute, which is operated by the Association of Universities for Research in Astronomy, Incorporated, under NASA contract NAS5-26555.
MagAO-X is funded by the NSF MRI program, award \#1625441.


\bibliography{report}   
\bibliographystyle{spiejour}   


\vspace{2ex}\noindent\textbf{Sebastiaan Y. Haffert} is a NASA Hubble Postdoctoral Fellow at the University of Arizona's Steward Observatory. His research focuses on high-spatial and high-spectral resolution instrumentation for exoplanet characterization.

\vspace{2ex}\noindent\textbf{Jared R. Males} is an Assistant Astronomer in the the University of Arizona's Steward Observatory and is the Principal Investigator of MagAO-X.  His research is focused on using high-contrast imaging to study extrasolar planets.

\vspace{2ex}\noindent\textbf{Joseph D. Long} is a Ph.D. student at the University of Arizona's Steward Observatory and a member of the MagAO-X instrument team researching exoplanet observation post-processing techniques.

\listoffigures
\listoftables

\end{spacing}
\end{document}